%% file: main.tex
\documentclass[3p]{elsarticle}

\usepackage[english]{babel}

\usepackage{amsmath}
\usepackage{amsthm}
\usepackage{amssymb}
\usepackage{graphicx}
\usepackage[colorlinks=true, allcolors=blue]{hyperref}
\usepackage{xcolor}
\usepackage{bbm}
\usepackage[capitalise]{cleveref}
\usepackage{caption}
\usepackage{subcaption}
\usepackage{enumitem}
\setlist{nosep}
\usepackage{titlesec}

\biboptions{sort&compress}
  
\newtheorem{theorem}{Theorem}[section]
\newtheorem{lemma}{Lemma}[section]
\newtheorem{definition}{Definition}[section]
\newtheorem{corollary}{Corollary}[section]

\makeatletter
\newtheorem*{rep@theorem}{\rep@title}
\newcommand{\newreptheorem}[2]{%
\newenvironment{rep#1}[1]{%
 \def\rep@title{#2 \ref{##1}}%
 \begin{rep@theorem}}%
 {\end{rep@theorem}}}
\makeatother

\newreptheorem{theorem}{Theorem}
\newreptheorem{lemma}{Lemma}

\newcommand{\Ep}{\mathbb{E}}
\newcommand{\Prob}{\mathbb{P}}
\newcommand{\msj}{\textnormal{MSJ}}
\newcommand{\ak}{{\textnormal{Ak}}}
\newcommand{\sat}{\textnormal{Sat}}
\newcommand{\sss}{\textnormal{SSS}}

\newcommand{\indic}{\mathbbm{1}}
\newcommand{\y}{\mathbb{Y}}
\newcommand{\ysmsj}{\mathbb{Y}^\msj}
\newcommand{\ysak}{\mathbb{Y}^\ak}
\newcommand{\yspi}{\mathbb{Y}^\pi}
\newcommand{\comp}{\lambda^*}
\newcommand{\comppi}{\lambda^*_\pi}
\newcommand{\pgood}{\text{pGood}}
\newcommand{\s}{\mathbb{S}}

\journal{IFIP Performance}

\begin{document}
\begin{frontmatter}
\title{The RESET and MARC Techniques, with Application to Multiserver-Job Analysis}
\author[cmu]{Isaac Grosof}
\author[cmu]{Yige Hong}
\author[cmu]{Mor Harchol-Balter}
\author[cmu]{Alan Scheller-Wolf}
\affiliation[cmu]{
organization={Carnegie Mellon University},
addressline={5000 Forbes Ave},
city={Pittsburgh},
postcode={15213},
state={PA},
country={USA\\},
email={\{igrosof, yigeh, harchol, awolf\}@andrew.cmu.edu}}
\begin{abstract}
    Multiserver-job (MSJ) systems, where jobs need to run concurrently across many servers, are increasingly common in practice.
    The default service ordering in many settings is First-Come First-Served (FCFS) service.
    Virtually all theoretical work on MSJ FCFS models focuses on characterizing the stability region,
    with almost nothing known about mean response time.
    
    We derive the first explicit characterization of mean response time in the MSJ FCFS system. Our formula characterizes mean response time up to an additive constant, which becomes negligible as arrival rate approaches throughput, and allows for general phase-type job durations.

    We derive our result by utilizing two key techniques: REduction to Saturated for Expected Time (RESET)
    and MArkovian Relative Completions (MARC).
    
    Using our novel RESET technique, we reduce the problem of characterizing mean response time in the MSJ FCFS system to an M/M/1 with Markovian service rate (MMSR). The Markov chain controlling the service rate is based on the saturated system, a simpler closed system which is far more analytically tractable.
    
    Unfortunately, the MMSR has no explicit characterization of mean response time.
    We therefore use our novel MARC technique to give the first explicit characterization of mean response time in the MMSR,
    again up to constant additive error.
    We specifically introduce the concept of ``relative completions,''
    which is the cornerstone of our MARC technique.
\end{abstract}

\end{frontmatter}
\section{Introduction}

Multiserver queueing theory predominantly emphasizes models in which each job utilizes only one server
(one-server-per-job models),
such as the M/G/k.
For decades, such models were popular in the study of computing systems,
where they provided a faithful reflection of the behavior of such systems
while remaining conducive to theoretical analysis.
However, one-server-per-job models no longer reflect the behavior of many modern computing systems.

\textbf{Multiserver jobs:}
In modern datacenters, such as those of Google, Amazon, and Microsoft,
each job now requests many servers (cores, processors, etc.),
which the job holds simultaneously.
A job's ``server need" refers to the number of servers requested by the job.
In Google's recently published trace of its ``Borg" computation cluster \cite{grosof_wcfs_2021,tirmazi_2020},
the server needs vary by a factor of 100,000 across jobs.
Throughout this paper, we will focus on this ``multiserver-job model" (MSJ),
in which each job requests some number of servers,
and concurrently occupies that many servers throughout its time in service (its ``duration'').

\textbf{FCFS service:}
We specifically study
the first-come first-served (FCFS) service ordering for the MSJ model,
a natural and practical policy that is the default in both cloud computing \cite{etsion_short,sliwko_taxonomy_2019,madni_performance_2017} and supercomputing \cite{feitelson_parallel_2004,jones_scheduling}.
Currently, little is known about FCFS service in MSJ models.

\textbf{Stability under FCFS:}
Even the stability region under FCFS scheduling is not generally understood.
Some papers characterize the stability region
under restrictive assumptions on the job duration distributions \cite{afanaseva_stability_2019,rumyantsev_2017,rumyantsev_three_2022,morozov_stability_2016,grosof_new_2023}.
A key technique in these papers is the
\emph{saturated system} approach \cite{foss_2004,baccelli_1995}.
The saturated system is a closed system in which completions trigger new arrivals,
so that the number of jobs in the system is always constant.
We are the first to use the saturated system
for analysis beyond
characterizing the stability region.

\textbf{Response time for FCFS:}
Even less is known about mean response time $\Ep[T]$ in MSJ FCFS systems:
The only MSJ FCFS system in which mean response time has been analytically characterized is the 
simpler case of 2 servers and exponentially distributed durations 
\cite{brill_queues_1984,fillippopoulos_mm2}.
Mean response time is much better understood under more complex scheduling policies such as ServerFilling and ServerFilling-SRPT \cite{grosof_wcfs_2021,grosof_optimal_2022}, but these policies require assumptions on both preemption and the server need distribution,
and do not capture current practices, which emphasize nonpreemptive policies.
Mean response time is also better understood
in MSJ FCFS scaling regimes, where the number of servers and the arrival rate both grow asymptotically
\cite{wang_zero_2021,hong_sharp_2022}.
We are the first to analyze MSJ FCFS mean response time
under a fixed number of servers.

\textbf{Why FCFS is hard to analyze:}
One source of difficulty in studying the FCFS policy is the lack of work conservation.
In simpler one-server-per-job models, a work-conservation property holds:
If enough jobs are present, no servers will be idle.
The same is true under the ServerFilling and ServerFilling-SRPT policies \cite{grosof_wcfs_2021},
which focus on the power-of-two server-need setting.
Each policy selects a subset of the jobs available,
and places jobs from that subset into service in largest-server-need-first order.
By doing so, and using the power-of-two assumption, these policies always fill all of the servers,
whenever sufficiently many jobs are present, thereby achieving work conservation.

Work conservation is key to the mean response time analysis of those systems,
as one can often reduce the analysis of response time
to the analysis of work.
In contrast, the multiserver-job model under FCFS service
is not work conserving:
a job must wait if it demands more servers than are currently available,
leaving those servers idle.

\textbf{First response time analysis:}
We derive the first characterization of mean response time in the MSJ FCFS system.
We allow any phase-type duration distribution, and any correlated distribution of server need and duration.
Our result holds at all loads up to an additive error,
which becomes negligible as the arrival rate $\lambda$
approaches $\comp$, the threshold of stability.

\begin{figure}
    \centering
    \includegraphics[width=\textwidth]{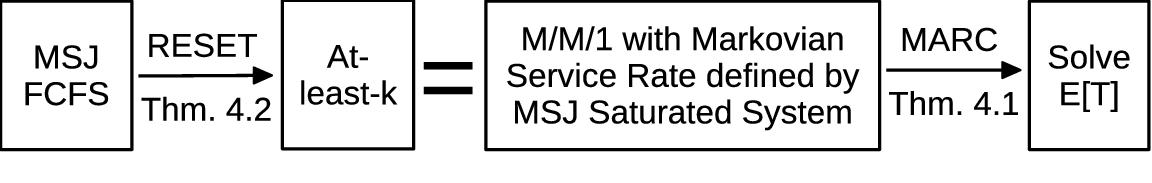}
    \caption{The structure of our main results: RESET (\cref{thm:msj-response-time}) and MARC (\cref{thm:mmsr-response-time}).}
    \label{fig:structure}
\end{figure}

\textbf{Proof structure:}
We illustrate the structure of our proof in \cref{fig:structure}.
We first use our RESET technique (REduction to Saturated for Expected Time)
to reduce from the MSJ FCFS system to the At-least-$k$ system (see \cref{sec:at-least-k}).
The At-least-$k$ system
is equivalent to a M/M/1 with Markovian service rate (MMSR) (see \cref{sec:mmsr}),
where the service rate is based on the saturated system.
By ``Markovian service rate'', we refer to a system in which the completion rate fluctuates over time,
driven by an external finite-state Markov chain.
We next use our MARC technique (MArkovian Relative Completions)
to prove \cref{thm:mmsr-response-time}, the first characterization of mean response time in the MMSR.

Both steps are novel, hard, and of independent interest.
We prove our MARC result first because it is a standalone result,
characterizing mean response time for any MMSR system up to an additive constant.
We then prove \cref{thm:msj-response-time}, our characterization of mean response time in the MSJ FCFS system, by layering our RESET technique on top of MARC.
\cref{thm:msj-response-time} characterizes mean response time in terms of several quantities that can be characterized explicitly and in closed form via a straightforward analysis of the saturated system.
We walk through a specific example of using our result
to explicitly characterize mean response time in \ref{app:calculate}.

\textbf{Breadth of the RESET technique:}
Our RESET technique is very broad, and applies to a variety of generalizations of the MSJ model
and beyond (See \cref{sec:beyond}).
For instance, RESET can handle cases where
a job's server need varies throughout its time in service,
and where the service rates at the servers can depend on the job.
Finally, we can analyze scheduling policies that are close to FCFS but allow limited reordering,
such as some backfilling policies.

\textbf{Breadth of the MARC technique:}
Our MARC technique is also very broad, and applies to any MMSR system. For example, we can handle systems in which machine breakdowns lead to reduced service rate, or where servers are taken away by higher-priority customers.

This paper is organized as follows:
\begin{itemize}
    \item \cref{sec:prior-work}: We discuss prior work on the MSJ model.
    \item \cref{sec:model-1}: We define the MSJ model, the MMSR, the saturated system, relative completions, and related concepts.
    \item \cref{sec:results-1}: We state our main results, and walk through an example of applying our results to a specific MSJ FCFS system.
    \item \cref{sec:marc}: We characterize mean response time in the MMSR using our MARC technique.
    \item \cref{sec:reset}: We build upon \cref{sec:marc} to characterize MSJ FCFS mean response time using our RESET technique.
    \item \cref{sec:beyond}: Our results apply to a very broad class of models which we call ``finite skip'' models, and which we define in this section.
    \item \cref{sec:empirical}: We empirically validate our theoretical results.
\end{itemize}

\section{Prior work}
\label{sec:prior-work}

The bulk of the prior work we discuss is in \cref{sec:prior-work-msj}, which focuses on specific results in the multiserver-job model. In \cref{sec:prior-work-saturated}, we briefly discuss prior work on the saturated system, an important tool in our analysis. Finally, in \cref{sec:prior-work-msr}, we discuss prior work on the M/M/1 with Markovian service rate.

\subsection{Multiserver-job model}
\label{sec:prior-work-msj}

Theoretical results in the multiserver-job model are limited.
We first discuss the primary setting of this paper:
a fixed number of servers and FCFS service.

\subsubsection{Fixed number of servers, FCFS service}
In this setting, most results focus on characterizing the stability region.
Rumyantsev and Morozov characterize stability for an MSJ system with an arbitrary distribution of server needs, where the duration distribution is exponential and independent of server need \cite{rumyantsev_2017}. This result can implicitly be seen as solving the saturated system, which has a product-form stationary distribution in this setting. A setting with two job classes, each with distinct server needs and exponential duration distributions has also been considered \cite{grosof_stability_2020,rumyantsev_stability_2020}. In this setting, the saturated system was also proven to have a product-form stationary distribution, which was also used to characterize the stability region. 

The only setting in which mean response time $\Ep[T]$ is known is in the case of $k=2$ servers and exponential duration independent of server need \cite{brill_queues_1984,fillippopoulos_mm2}. In this setting, the exact stationary distribution is known. Mean response time is open in all other settings, including whenever $k>2$.

\subsubsection{Advanced scheduling policies}

More advanced scheduling policies for the MSJ system have been investigated, in order to analyze and optimize the stability region and mean response time.

The MaxWeight policy
was proven to achieve optimal stability region in the MSJ setting \cite{maguluri_scheduling_2014}.
However, its implementation requires solving an NP-hard optimization problem upon every transition,
and it performs frequent preemption.
It is also too complex for response time analysis to be tractable.
The Randomized Timers policy achieves optimal throughput with no preemption \cite{ghaderi_randomized_2016,psychas_randomized_2018},
but has very poor empirical mean response time, and no response time analysis.

In some settings, it is possible for a scheduling policy to ensure that all servers are busy whenever there is enough work in the system, which we call ``work conservation."
Work conservation enables the optimal stability region to be achieved
and mean response time to be characterized.
Two examples are ServerFilling and ServerFilling-SRPT scheduling policies \cite{grosof_wcfs_2021,grosof_optimal_2022}.
However, the work-conservation-based techniques used in these papers cannot be used to analyze non-work-conserving policies such as FCFS.

\subsubsection{Scaling number of servers}

The MSJ FCFS model has also been studied in settings where the number of servers, the arrival rate, and the server need distribution all grow in unison to infinity.
Analogues of the Halfin-Whitt and non-diminishing-slowdown regimes have been established, proving bounds on the probability of queueing and mean waiting time \cite{hong_sharp_2022,wang_zero_2021}.
These results focus on settings where an \emph{approximate} work conservation property holds, and there is enough excess capacity that this approximate work conservation is sufficient to determine the first-order behavior of the system.
These results do not apply to the $\lambda \to \lambda^*$ limit.

\subsection{Prior work on the saturated system}
\label{sec:prior-work-saturated}

The \emph{saturated system} is a queueing system which is used as analysis tool to understand the behavior of an underlying non-saturated queueing system
\cite{baccelli_1995,foss_2004}.
Baccelli and Foss state that it is a ``folk theorem" that the threshold of the stability region of the original open queueing system is equivalent to the completion rate of the saturated system: If the completion rate of the saturated system is $\mu$, then the original system is stable for arrival rate $\lambda$ if and only if $\lambda < \lambda^* = \mu$ \cite{baccelli_1995}. Baccelli and Foss give sufficient conditions for this folk theorem, known as the ``saturation rule," to hold rigorously.
These conditions are mild, and are easily shown to hold for the MSJ FCFS system.
The strongest stability results in the MSJ FCFS system have either been proven by characterizing the steady state of the saturated system, or are equivalent to such a characterization \cite{rumyantsev_2017,grosof_stability_2020,grosof_new_2023}.

Our novel contribution is characterizing the \emph{mean response time} behavior of an original system
by reducing its analysis to the analysis of a saturated system.
All previous uses of the saturated system focused on characterizing stability.
Specifically, our main theorem, \cref{thm:msj-response-time}, characterizes mean response time in terms of $\Delta_{\sat}(y), \comp,$ and $Y_d^\sat$.
These functions and random variables are specific to the saturated system.
They are defined in \cref{sec:model-1},
and can be calculated in closed-form by analyzing the saturated system,
as we walk through in \ref{app:calculate}.

\subsection{M/M/1 with Markovian Service Rate}
\label{sec:prior-work-msr}

The M/M/1 with Markovian service rate (MMSR) has been extensively studied since the 50's, often alongside Markovian arrival rates \cite{clarke_waiting_1956,neuts_single_1966,massey_asymptotic_1985,knessl_exact_2002,gupta_fundamental_2006,delasay_modeling_2016}.
A variety of mathematical tools have been applied to the MMSR, including generating function methods, matrix-analytic and matrix-geometric methods, and spectral expansion methods \cite{clarke_waiting_1956,neuts_single_1966,lucantoni_some_1994,delasay_modeling_2016}.
However, these methods primarily result in \emph{numerical results}, rather than theoretical insights \cite{mitrani_spectral_1995,delasay_modeling_2016}.

More is known for special cases of the MMSR system \cite{perel_queues_2008,doroudi_stochastic_2016}.
For instance, the case where arrival rates alternate between a high and low completion rate at some frequency 
has received specific study. In this case, the generating function can be explicitly solved as the root of a cubic equation \cite{yechiali_queuing_1971}, but the resulting expression is too complex for analytical insights.
In this simplified setting, scaling results \cite{vesilo_scaling_2022,newell_queuesI_1968,newell_queuesII_1968,newell_queuesIII_1968} and monotonicity results \cite{gupta_fundamental_2006} have been derived, but those results do not extend to more complex MMSR systems.

By contrast, our MARC technique provides the first explicit characterization of mean response time for the general MMSR system, up to an additive constant.

\subsection{Drift method and MARC}


The drift method is a popular method for steady-state analysis of queueing models (see, e.g., \cite{eryilmaz_drift_2012,Maguluri_drift_16,hong_sharp_2022,wang_zero_2021}). 
In the drift method, one takes a suitable \emph{test function} (also known as a Lyapunov function) of the system state and computes its instantaneous rate of change starting from each state under the transition dynamics, which is called the drift. The drift can be formally calculated using the \emph{instantaneous generator}, defined in \cref{sec:generator}.
One then utilizes the fact that the drift of any test function has zero steady-state expectation (\Cref{lem:drift-lemma}) to characterize system behavior in steady state, through metrics such as mean queue length. Through more specialized choices of test function, stronger results such as State Space Collapse can also be proven.

In prior work which analyzes the mean queue length,
the test function is usually a quadratic function of the queue length.
For instance, when analyzing the MaxWeight policy in the switch setting,
an appropriate test function is $\sum_i q_i^2$,
where $q_i$ is the number of jobs present of each class $i$ \cite{srikant_communication_2013}.
For such a test function to provide useful information about the expected queue length,
the system must achieve a constant work completion rate whenever
there are enough jobs in the system.
This constant work completion rate ensures that the test function's drift
depends linearly on the queue length, allowing the mean queue length to be characterized.
However, in our MSJ system, the work completion rate is variable regardless of the number of jobs in the system, because servers may always be left empty if a job in the queue requires more servers than are available. As a result, the standard test functions for the drift method do not provide useful information about the MSJ system.

Our innovation is to construct a novel test function that combines the queue length $q$ and a new quantity called \emph{relative completions}, defined in \Cref{sec:delta}.
Our use of relative completions allows us to ensure that the test functions $f_\Delta$ and $f_{\Delta}^\msj$, defined in \cref{def:f-delta,def:f-delta-msj}, have drift which depend linearly on the queue length.
As a result, we can apply the drift method with our novel test functions
to characterize mean queue length in the MSJ system, and hence characterize mean response time.

We call this technique the MArkovian Relative Completions (MARC) technique: using relative completions to define a test function for the drift method, to apply the drift method to systems with variable work-completion rate.

\section{Model}
\label{sec:model-1}

\begin{table}
    \centering
    \begin{tabular}{c | c | c}
        Abbreviation & Meaning & Definition \\
        \hline
        MSJ & Multiserver-job & \cref{sec:model-msj} \\
        FCFS & First-come first-served & \cref{sec:model-msj} \\
        MMSR & M/M/1 with Markovian service rate & \cref{sec:mmsr} \\
        Ak & At-least-$k$ system & \cref{sec:at-least-k} \\
        Sat & Saturated system & \cref{sec:saturated} \\
        SSS & Simplified saturated system & \cref{sec:simplified}, \ref{app:sss} \\
        MARC & Markovian relative completions & \cref{sec:marc} \\
        RESET & Reduction to saturated for expected time & \cref{sec:reset}
    \end{tabular}
    \caption{Table of abbreviations}
    \label{tbl:abbrev}
\end{table}

In this section, we introduce five queueing models: the multiserver-job (MSJ) model, the M/M/1 with Markovian service rate (MMSR), the At-least-$k$ (Ak) model, the saturated system, and the simplified saturated system (SSS).
The MSJ is the main focus of this paper.
Our RESET technique reduces its analysis to analyzing the Ak system.
The Ak system is equivalent to a MMSR system whose completion process is controlled by the saturated system. Our MARC technique allows us to analyze this MMSR system. The SSS is a simpler equivalent of the saturated system.
We also introduce the concepts of relative completions and the generator approach, which are key to our analysis.

\cref{tbl:abbrev} describes each of the abbreviations used in this paper.

\subsection{Multiserver-job Model}\label{sec:model-msj}

The MSJ model is a queueing model in which each job requests an integer number of servers,
the \emph{server need},
for some duration of time, the \emph{service duration}.
Each job requires concurrent service on all of its servers throughout its duration.
Let $k$ denote the total number of servers in the system.

We assume that each job's server need and service duration are drawn i.i.d. from some joint distribution.
The duration distribution is phase type,
and it may depend on the job's server need.
This assumption can likely be generalized, which we leave to future work.
We assume a Poisson($\lambda)$ arrival process.

We focus on the first-come first-served (FCFS) service discipline.
Our RESET technique also applies to many other scheduling policies,
as we discuss in \cref{sec:beyond}.
Under FCFS,
jobs are placed into service, one by one, in arrival order,
as long as the total server need of the jobs in service is at most $k$.
If a job is reached whose server need would push the total over $k$,
that job does not receive service until sufficient completions occur.
We consider head-of-the-line blocking, so no subsequent jobs in arrival order receive service.
It has been shown that in the MSJ FCFS setting, there exists a threshold $\lambda^*$,
such that the system is stable if and only if $\lambda < \lambda^*$ \cite{baccelli_1995, foss_2004}.
We assume that $\lambda < \lambda^*$.

Note that the only jobs eligible for service are the $k$ oldest jobs in arrival order.
We conceptually divide the system into two parts:
the \emph{front} and the \emph{back}. When the total number of jobs in the system is at least $k$, the front consists of the $k$-oldest jobs in the arrival order; otherwise, the front consists of all jobs in the system. The back consists of all jobs that are not in the front. 
Note that all of the jobs which are in service must be in the front, because at most $k$ jobs can be in service at a time, and service proceeds in strict FCFS order.
The front may also contain some jobs which are not in service, whenever less than $k$ jobs are in service.
All of the jobs in the back are not in service.

\subsection{M/M/1 with Markovian Service Rate}
\label{sec:mmsr}

The MMSR-$\pi$ system
is a queueing system where jobs arrive to the system according to a Poisson process, and complete at a variable rate,
where the completions are determined by the transitions of a finite-state Markov chain $\pi$.
We refer to $\pi$ as the ``service process''.
When a job arrives, it stays in the queue until it reaches the head of the line,
entering service.
The job then completes when $\pi$ next undergoes a transition associated with a completion.
Jobs are identical until they reach service.
The service process $\pi$ is unaffected by the number of jobs in the queue.

\subsection{At-least-$k$ System}
\label{sec:at-least-k}

To connect the MSJ FCFS and MMSR systems, we define two systems: the ``At-least-$k$" (Ak) system,
and the ``saturated system'' in \cref{sec:saturated}.
The Ak model mimics the MSJ model,
except that the Ak system always has at least $k$ jobs present. 
Specifically, in addition to the primary Poisson($\lambda$) arrival process, 
whenever there are exactly $k$ jobs in the system,
and a job completes, a new job immediately arrives.
The server need and service duration of this job are sampled i.i.d. from the same distribution
as the primary arrivals.
Due to these extra arrivals, the front of the Ak system always has exactly $k$ jobs present.

Intuitively, the Ak system should have about $k$ more jobs present in steady state than the MSJ system.
We thus expect the Ak and MSJ systems to have the same asymptotic mean response time, up to an $O_\lambda(1)$ term.
We make this intuition rigorous by using our RESET technique to prove \cref{thm:msj-response-time}.

\subsection{Running Example}
\label{sec:running}

Throughout this section, we will use a running example to clarify notation and concepts.
Consider a MSJ setting with $k=2$ servers, and two classes of jobs: $2/3$ of jobs have server need 1 and duration $Exp(1)$, and the other $1/3$ of jobs have server need $2$ and duration $Exp(1/2)$.

\subsection{Saturated System}
\label{sec:saturated}

The saturated system is a closed multiserver-job system, where completions trigger new arrivals.\footnote{\citet{baccelli_1995} consider a system with infinitely many jobs
not in service, which is equivalent to our closed system.}
Jobs are served according to the same FCFS service discipline.
There are always exactly $k$ jobs in the system.
Whenever a job completes, a new job with i.i.d. server need and service duration is sampled.
The state descriptor is just an ordered list of exactly $k$ jobs.

In our running example with $k=2$ servers, the state space of the saturated system consists of all orderings of $2$ jobs:
\begin{align*}
    \y^\sat = \{[1,1], [1, 2], [2, 1], [2,2]\}.
\end{align*}
The leftmost entry in each of the lists is the oldest job in FCFS order. In state $[1,2]$, a 1-server job is in service and a 2-server job is not in service, while in state $[2,1]$, a 2-server job is in service and a 1-server job is not in service.

\subsection{Equivalence between MMSR-Sat and At-least-$k$}
\label{sec:equivalence}

Now we are ready to connect the MMSR and At-least-$k$ (Ak) systems.
Consider the subsystem consisting only of the front of the Ak system, i.e., the $k$ oldest jobs in the Ak system.
This subsystem is stochastically identical to the saturated system.
Whenever a job completes at the front of the Ak system,
a new job enters the front, either from the back (i.e. the jobs not in the front) or from the auxiliary arrival process, if the back is empty.
This matches the saturated system's completion-triggered arrival process.

As a result, the Ak system is stochastically equal to an MMSR-$\pi$ system
whose service process $\pi$ is identical to the saturated system.
We refer to this system as the ``MMSR-Sat'' system.
To clarify this equivalency, assume the Ak system starts in a certain front state $y$ with an empty back.
Then equivalently the MMSR-Sat system starts empty, with its service process in state $y$.
If a job in the Ak system completes its service,
a new job is generated,
and the same transition occurs in the service process in the MMSR-Sat system.
Similarly, assume a job arrives to the Ak system and enters the back.
At the same time, a job arrives in the MMSR-Sat system and enters the queue.
Through this mapping, the two systems are sample-path equivalent.

The above arguments are summarized in \cref{lem:ak-equiv-mmsr-sat} below.
\begin{lemma}
\label{lem:ak-equiv-mmsr-sat}
    There exists a coupling under which the front of the Ak system is identical to the Sat system, and the back of the Ak system is identical to the queue of the MMSR-Sat system.
\end{lemma}
\subsection{Notation}

\textbf{MSJ system state:}
A state of the MSJ system consists of a front state, $y^\msj$,
and a number of jobs in the back $q^\msj$.
A job state consists of a server need
and a phase of its phase-type duration.
The front state $y^\msj$ is a list of up to $k$ job states.
If $q^\msj > 0$, then $y^\msj$ must consist of exactly $k$ job states,
while if $q^\msj = 0$, $y^\msj$ may consist of anywhere from $0$ to $k$ job states.
Let $\ysmsj$ denote the set of all possible front states $y^\msj$ of the MSJ system.
For instance, in our running example, $\ysmsj=\{[], [1], [2], [1,1], [1, 2], [2, 1], [2,2]\}$.
Note that in the first three states, the back must be empty, so $q^\msj$ must equal $0$.

\textbf{MMSR system state:}
In the MMSR system,
let $\pi$ denote the Markov chain that modulates the service rate.
As a superscript, it signifies ``the MMSR system controlled by the Markov chain $\pi$.''
A state of the MMSR-$\pi$ system consists of a pair $(q^\pi, y^\pi)$.
The queue length $q^\pi$ is a nonnegative integer.
The state $y^\pi$ is a state of the service process $\pi$, and $\yspi$ is the state space of $\pi$.

Because the MMSR-$\sat$ system is stochastically equal to the Ak system,
with the MMSR-$\sat$ system's queue length equal to the Ak system's back length,
we use the superscripts $^\sat$ and $^\ak$ interchangeably.
A state of the Ak system is a pair $(q^\ak, y^\ak)$.
In contrast to the MSJ system, $y^\ak$ always consists of exactly $k$ job states.
In particular, $\ysak \subset \ysmsj$.

\textbf{MMSR service process:}
When the service process $\pi$ transitions from state $y$ to $y'$, there are two possibilities:
Either a completion occurs, which we write as $a=1$,
or no completion occurs, which we write as $a=0$.
We therefore define $\mu^\pi_{y,y',a}$ to denote
the system's transition rate from front state $y$ to front state $y'$,
accompanied by $a$ completions, where $a \in \{0, 1\}$.
For instance, in our running example $\mu^\sat_{[1,1], [1,2],1} = 2/3$.
Let the total completion rate from state $y$
be denoted by $\mu^\pi_{y,\cdot,1} = \sum_{y'} \mu^\pi_{y, y', 1}$.
For instance, in our running example $\mu^\sat_{[1,1],\cdot,1}=2$.

\textbf{MSJ service transitions:}
Let $\mu^\msj_{y,y',a,b}$ denote a transition rate in the Multiserver-job system,
where
$y,y',$ and $a$ have the same meaning as in $\mu^\ak_{y,y',a}$.
Let $b = \mathbbm{1}_{q>0}$
denote whether this transition is associated with an empty back ($b=0$), or an occupied back ($b=1$).
Note that if $y \not\in \ysak$, then $b=0$ for all nonzero $\mu^\msj_{y,y',a,b}$,
while if $y \in \ysak$, then both values of $b$ are possible.
Note that $\forall y \in \ysak, \mu^\msj_{y,y',a,1}=\mu^\ak_{y,y',a}$.

If a job arrives to the MSJ system and finds that the front state $y$ has fewer than $k$ jobs
($y \not\in \ysak$), a fresh job state is sampled and appended to $y$.
Let $S$ be a random variable denoting a fresh job state,
let $i$ be a particular fresh job state,
let $p_i$ be the probability $\Prob(S = i)$,
and let $y \cdot i$ be the new front state with
a job in state $i$ appended.
For instance, in the running example, $p_1=2/3, p_2=1/3$.

\textbf{Steady-state notation:}
We will study the time-average steady states of each of these systems,
which we write $(Q^\msj, Y^\msj)$, $(Q^\pi, Y^\pi)$, etc.
Let $Y^\pi_d$ denote the departure-average steady state of the MMSR service process $\pi$:
the steady-state distribution of the embedded DTMC
which samples states after each departure from $\pi$.

Let $X^\pi$ denote the long-term throughput of the service process $\pi$.
Let $\lambda^*_\pi$ denote the threshold of the stability region of the MMSR-$\pi$ system.
The MMSR-$\pi$ system is stable if and only if $\lambda < \lambda^*$.
Note that $X^\pi = \lambda^*_\pi$ by prior results relating the saturated system to the stability region of the original system \cite{baccelli_1995,foss_2004}.
In particular, $X^\sat = \lambda^*_\sat= \lambda^*$,
where $\lambda^*$ denotes the threshold of the stability region of the MSJ FCFS system.
We will typically write $\lambda^*$ to avoid confusion between $X^\sat$ and a random variable.

A concrete example of this notation is provided in \cref{sec:example}.

\subsection{Relative completions}
\label{sec:delta}

Key to our MARC technique is the novel idea of \emph{relative completions},
which we define for a general MMSR-$\pi$ system.
Let $y_1$ and $y_2$ be two states of the service process $\pi$.
The difference in relative completions between two states $y_1$ and $y_2$
is the long-term difference in expected completions between
an instance of the service process starting in state $y_1$ and one starting in $y_2$.
Specifically, let $C_\pi(y, t)$ denote the number of completions up to time $t$ of the
service process of $\pi$
initialized in state $y$ at time $t=0$.
Then let $\Delta_\pi(y_1,y_2)$ denote the relative completions between states $y_1$ and $y_2$:
\begin{align*}
    \Delta_\pi(y_1,y_2) = \lim_{t \to \infty} \Ep[C_\pi(y_1,t) - C_\pi(y_2,t)].
\end{align*}
We prove that $\Delta_\pi(y_1,y_2)$ always exists and is always finite
in \cref{lem:delta-exists}.
We also allow $y_1$ and/or $y_2$ to be distributions over states, rather than single states. Specifically,
we will often focus on the case where $y_2$, rather than being a single state,
is the steady state distribution $Y^\pi$.
In this case, note that $\Ep[C_\pi(Y^\pi,t)]=X^\pi t=\comp_\pi t$.
When it is clear from context, we write $\Delta_\pi(y)$ to denote $\Delta_\pi(y,Y^\pi)$.
The relative completions formula for this case simplifies:
\begin{align}
    \label{eq:simple-delta}
    \Delta_\pi(y) = \Delta_\pi(y,Y^\pi) = \lim_{t \to \infty} \Ep[C_\pi(y,t)] - \comp_\pi t.
\end{align}

The relative completions function $\Delta_\pi(y)$
can be seen as the relative value of a given state $y$ under a
Markov reward process whose state is a state of the service process $\pi$
and whose reward is the instantaneous completion rate in a given state $y$.

\subsection{Generator}
\label{sec:generator}

We also make use of the \emph{instantaneous generator} of each of our queueing systems,
which is the stochastic equivalent of the derivative operator.
The instantaneous generator is an operator which takes a function from system states to real values,
and returns a function from system states to real values.
The latter function is known as the \emph{drift} of the original function.

The generator operator is specific to a given Markov chain.
Let $\eta$ be a Markov chain,
and let $G^\eta$ denote the generator operator for $\eta$, which is defined as follows:

For any real-valued function of the state of $\eta$, $f(q, y)$, 
\begin{align*}
    G^\eta \circ f(q, y) := \lim_{t \to 0} \frac{1}{t} \Ep[f(Q^\eta(t), Y^\eta(t)) - f(q, y) | Q^\eta(0) = q, Y^\eta(0)=y].
\end{align*}

Importantly, the expected value of the generator in steady state is zero:
\begin{lemma}\label{lem:drift-lemma}
    Let $f$ be a real-valued function of the state of a Markov chain $\eta$. Assume that the transition rates of the Markov chain $\eta$ are uniformly bounded, and $\Ep[f(Q^\eta, Y^\eta)] < \infty.$
    Then 
    \begin{equation}\label{eq:generator-steady-zero}
        \Ep_{(q,y) \sim (Q^\eta, Y^\eta)} [G^\eta \circ f(q, y)] = 0.
    \end{equation}
\end{lemma}
\begin{proof}
Follows from \cite[Proposition~3]{glynn_bounding_2008}.
Discussion deferred to \ref{app:basic-results}.
\end{proof}
We show in \ref{app:basic-results} that
\eqref{eq:generator-steady-zero} holds for the MSJ, MMSR, At-least-$k$, and Saturated systems, for any $f(q, y)$ with polynomial dependence on $q$.

\subsection{Asymptotic notation}

We use the notation $O_\lambda(f(\lambda))$
to represent a function $g(\lambda)$ such that
\begin{align*}
    \exists \text{ a constant } M \text{ such that } |g(\lambda)| \le M|f(\lambda)| \quad \forall\lambda, 0 < \lambda < \lambda^*.
\end{align*}

\subsection{Simplified saturated system}
\label{sec:simplified}

While the saturated system is a finite-state system,
it can have a very large number of possible states.
However, many of the states have identical behavior, and can be combined to reduce the state space.
For instance, in our running example, the states $[2,1]$ and $[2,2]$ are nearly identical:
in both states just a 2-server job is in service.
We therefore simplify the system by combining the two states into the state $[2]$,
and delaying sampling the next job until needed.

We refer to the resulting system as the ``simplified saturated system'' (SSS), in contrast to the original saturated system, which is the focus of the bulk of this paper. SSS is equivalent to the original saturated system, in the sense of \ref{lem:sss-coupling} stated below. 
\begin{lemma}
    \label{lem:sss-coupling}
    There exists a coupling under which the main saturated system
    and simplified saturated system
    have identical completions.
\end{lemma}

The full definition of the SSS, and the proof of the equivalence of SSS to the original saturated system, 
are in \ref{app:sss}.

The reduction in state space from the SSS can be dramatic. For instance, consider a system where $k=30$, jobs have server needs 3 or 10, and jobs have exponential duration.
The original saturated system has $2^{30}$ states, while the SSS has just $13$ states.
We discuss this reduction further in \ref{app:sss}.

\section{Results}
\label{sec:results-1}

In this paper, we give the first analysis of mean response time
in the MSJ FCFS system.
To do so, we reduce the problem to the analysis of mean response time
in an M/M/1 with Markovian service rate (MMSR)
in which the saturated system controls the service process (i.e. the At-least-$k$ system).
We call this reduction the RESET technique.
Before applying the RESET technique, we start by analyzing the general MMSR-$\pi$ system.

We prove the first explicit characterization of mean response time
in the MMSR.
To do so, we use our MARC technique,
which is based on the novel concept of \emph{relative completions}
(See \cref{sec:delta}).

\begin{theorem}[Mean response time asymptotics of MMSR systems]
    \label{thm:mmsr-response-time}
    In the MMSR-$\pi$ system, the expected response time in steady state satisfies
    \begin{align}
        \label{eq:mmsr-result}
        \Ep[T^\pi] = \frac{1}{\lambda^*_\pi} \frac{1+\Delta_\pi(Y^\pi_d, Y^\pi)}{1-\lambda/\lambda^*_\pi}+O_\lambda(1),
    \end{align}
    where $\Delta_\pi$ is the relative completions function defined in \cref{sec:delta}:
    \begin{align*}
        \Delta_\pi(Y^\pi_d, Y^\pi) := \lim_{t\to\infty} \Ep[C_\pi(Y_d^\pi, t)] - \comppi t.
    \end{align*}
\end{theorem}

To understand \eqref{eq:mmsr-result}, first note that the dominant term has order $\Theta(\frac{1}{1-\lambda/\lambda^*_\pi})$.
This is the equivalent of the $\Theta(\frac{1}{1-\rho})$ behavior seen in simpler systems such as the M/G/1/FCFS.
Next, to understand the numerator, examine the $\Delta_\pi(Y^\pi_d, Y^\pi)$ term.
$\Delta_\pi$, the relative completions function,
smooths out the irregularities in completion times,
so that the function $q-\Delta_\pi(y)$ has a constant negative drift.
$\Delta_\pi$ is the analog of the remaining size of the job in service in the M/G/1.
When a generic job arrives, it sees a time-average state of the service process, namely $Y^\pi$.
When it departs, it leaves behind a departure-average state of the service process,
namely $Y^\pi_d$.
The difference in relative completions between these states captures the asymptotic behavior of mean response time.
The overall numerator, $1+\Delta_\pi(Y^\pi_d, Y^\pi)$,
is analogous to the $\Ep[S_e]$ term in
the M/G/1/FCFS mean response time formula.
We walk through calculating $\Delta_\pi(y), \comppi,$ and $Y_d^\pi$ explicitly and in closed-form in \ref{app:calculate}.

Now that we have characterized the mean response time of the MMSR system,
we can use this result to characterize the MSJ FCFS system.
With our RESET technique, we show that the MSJ FCFS system has the same mean response time,
up to an $O_\lambda(1)$ term,
as the MMSR system whose service rate is controlled by the saturated system,
or equivalently the At-least-$k$ system.

\begin{theorem}[Mean response time asymptotics of MSJ systems]\label{thm:msj-response-time}
    In the multiserver\nobreakdash-job system, the expected response time in steady state satisfies
    \begin{align}
        \label{eq:msj-response-time}
        \Ep[T^\msj] = \frac{1}{\lambda^*} \frac{1+\Delta_\sat(Y^\sat_d, Y^\sat)}{1-\lambda/\lambda^*}+O_\lambda(1).
    \end{align}
\end{theorem}
Empirically, the $O_\lambda(1)$ term is very small, as seen in \cref{fig:general_response} in \cref{sec:empirical}.
To clarify the meaning of the $O_\lambda(1)$ term in \cref{thm:msj-response-time},
let us restate the theorem explicitly:
\begin{theorem}[Restatment of \cref{thm:msj-response-time}]
    In the multiserver\nobreakdash-job system,
    for any joint duration and server need distribution
    and for any number of servers $k$,
    there exist constants $c_\ell$ and $c_h$ such that
    for all arrival rates $\lambda < \comp$,
    \begin{align*}
        \frac{1}{\lambda^*} \frac{1+\Delta_\sat(Y^\sat_d, Y^\sat)}{1-\lambda/\lambda^*}+c_\ell \le \Ep[T^\msj] \le \frac{1}{\lambda^*} \frac{1+\Delta_\sat(Y^\sat_d, Y^\sat)}{1-\lambda/\lambda^*}+c_h.
    \end{align*}
\end{theorem}

Rather than calculating $\Delta_\sat(Y^\sat_d,Y^\sat)$ in \cref{thm:msj-response-time}, we can calculate the equivalent value in the simplified saturated system (SSS) (due to \cref{lem:sss-coupling}).
Define $\Delta_\sss, Y^\sss_d,$ and $Y^\sss$  analogously to the primary saturated system.

\begin{corollary}
    \label{cor:msj-sss-response-time}
    In the MSJ FCFS model,
    \begin{align*}
        \Ep[T^\msj] &= \frac{1}{\lambda^*} \frac{1 + \Delta_\sss(Y^\sss_d, Y^\sss)}{1-\lambda/\lambda^*} + O_\lambda(1).
    \end{align*}
\end{corollary}
\cref{cor:msj-sss-response-time} follows from \cref{thm:msj-response-time} because $\Delta_\sat(y_1, y_2)$ is defined based on the completion times in the primary saturated system, and by \cref{lem:sss-coupling},
the SSS can be coupled to have the same completion times as the primary saturated system.

The quantities $\Delta_{\sss}(y),\comp,$ and $Y_d^\sss$ can be calculated explicitly and in closed-form for any given parameterized distribution of server need and job duration, and any number of servers $k$,
giving an explicit closed-form bound on mean response time.
We walk through this calculation in \ref{app:calculate},
and give the explicit closed-form expressions for a 2-server setting in \ref{app:calculate-k2}, to demonstrate the technique.

\subsection{Example for demonstration}
\label{sec:example}

We now demonstrate applying \cref{thm:msj-response-time,cor:msj-sss-response-time} to characterize the asymptotic mean response time of our running example from \cref{sec:running}. See \ref{app:calculate} for a more extensive example,
handling a setting with parameterized completion rates and arrival probabilities.

We start with the MSJ system. First, we convert to the Ak system,
whose front has state space $\ysak=\{[1,1], [1,2], [2,1], [2,2]\}$.
By the RESET technique, this only increases mean response time by $O_\lambda(1)$.
By \cref{lem:ak-equiv-mmsr-sat}, the Ak system is identical to a MMSR-Sat system.
By \cref{lem:sss-coupling}, the $\sat$ system is equivalent to Simplified Saturated System (SSS), which has
state space $\mathbbm{Y}^\sss = \{ [1,1], [1,2], [2] \}.$

For the rest of this section, we focus on the SSS, leaving the superscript implicit.
Transitions between these states only happen as a result of completions,
leading to the following transition rates:
\begin{align*}
    \mu_{[1,1], [1, 1], 1} &= 2 \cdot \frac{2}{3} = \frac{4}{3}, \quad 
    \mu_{[1,1], [1, 2], 1} = 2 \cdot \frac{1}{3} = \frac{2}{3}, \qquad \quad
    \mu_{[1,2], [2],1} = 1, \\
    \mu_{[2],[1,1],1} &= \frac{1}{2} \frac{2}{3} \frac{2}{3} = \frac{2}{9}, \quad
    \mu_{[2],[1,2],1} = \frac{1}{2} \frac{2}{3} \frac{1}{3} = \frac{1}{9}, \quad
    \mu_{[2], [2], 1} = \frac{1}{2} \frac{1}{3} = \frac{1}{6}.
\end{align*}

Now, we can calculate the steady states $Y^\sss$ and $Y_d^\sss$
of the SSS's CTMC and DTMC respectively,
and calculate the throughput $X^\sss = X^\sat = \comp$.
The vectors are in the order $\{ [1,1], [1,2], [2] \}$:
\begin{align*}
    Y &= \Big[\frac{1}{5}, \frac{1}{5}, \frac{3}{5} \Big], \quad
    Y^d = \Big[ \frac{4}{9}, \frac{2}{9}, \frac{1}{3} \Big], \quad
    X^\sss = X^\sat = \comp = \frac{9}{10}.
\end{align*}

Now, we can solve for $\Delta(y)$, defined in \eqref{eq:simple-delta}.
To do so, we split up the completions $\Ep[C(y,t)]$ into the time until the first completion,
and the time after the first completion.
For example, starting in state $y=[1,1]$,
the first completion takes an expected $\frac{1}{2}$ second, during which 1 completion occurs,
compared to the long-term average rate $\frac{1}{2}\comp = \frac{9}{20}$ completions.
The system then transitions to a new state, with corresponding $\Delta(y)$.
This gives rise to the following equation:
\begin{align*}
    \Delta([1,1]) = 1 - \frac{9}{20} + \frac{2}{3} \Delta([1,1]) + \frac{1}{3} \Delta([1,2]).
\end{align*}

We use the same process to derive a system of equations that uniquely determines $\Delta(y)$,
given in \cref{cor:forward-delta}.
We solve for $\Delta(y)$ for each state $y$:
\begin{align}
    \label{eq:relative}
    \Delta([1,1]) &= 1.38, \quad \Delta([1,2]) = -0.27, \quad \Delta([2]) = -0.37.
\end{align}
All decimals are exact.
We can then average over the distribution $Y^d$
to find that $\Delta(Y^d) = 0.43$.
Recall that $\Delta(Y^d)$ is just shorthand for $\Delta(Y^d, Y).$

We can therefore apply \cref{thm:msj-response-time,cor:msj-sss-response-time}
to characterize the asymptotic mean response time of the original system:
\begin{align*}
    \Ep[T^\msj] = \frac{10}{9} \frac{1.43}{1 - \frac{\lambda}{9/10}} + O_\lambda(1).
\end{align*}
\section{MARC Proofs}
\label{sec:marc}

We start by analyzing the M/M/1 with Markovian service rate $\pi$ (MMSR-$\pi$). 
Our main result in this section is the proof of \cref{thm:mmsr-response-time},
a characterization of the asymptotic mean response time of the MMSR-$\pi$ system.

The main challenge is choosing an appropriate test function $f(q,y)$,
to leverage \eqref{eq:generator-steady-zero},
the fact that $\Ep[G^\pi \circ f(Q^\pi, Y^\pi)]=0$,
to give an expression for $\Ep[Q^\pi]$.
To gain information about $\Ep[Q^\pi]$ via this approach,
it is natural to choose a function $f$ which is quadratic in $q$,
because $G^\pi$ is effectively a derivative.
However, if we choose $f_1(q,y)=\frac{1}{2}q^2$,
the expression $G^\pi \circ f_1(q,y)$ will have cross-terms
in which both $q$ and $y$ appear, preventing further progress.

Instead, our key idea is to use relative completions $\Delta_\pi$ in our test function:

\begin{definition}
    \label{def:f-delta}
    Let $f_\Delta^\pi(q,y) = \frac{1}{2}(q-\Delta_\pi(y))^2$.
\end{definition}
The $\Delta_\pi(y)$ term smooths out the fluctuations in the system's service rate,
so that the quantity $q-\Delta_\pi(y)$ has a constant drift of $-\comppi$
whenever $q>0$.

This choice of test function
ensures that $G^\pi \circ f_\Delta^\pi(q,y)$ separates into a linear term dependent only on $q$ and a term dependent only on $y$.
The separation allows us to characterize $\Ep[Q^\pi],$ and hence $\Ep[T^\pi],$
in \cref{thm:mmsr-response-time}.






Let $u=\mathbbm{1}\{q =0 \land a=1\}$
denote the unused service
caused by a given transition. Only completion transitions ($a=1$) can cause unused service.

We start by decomposing $G^\pi \circ f_\Delta^\pi(q, y)$,
into a term linearly dependent on $q$,
and terms dependent only on $y, a,$ and $u$:
\begin{lemma}
    \label{lem:generator-f}
    For any state $(q, y)$ of the MMSR-$\pi$ system,
    \begin{align}
        G^\pi \circ f_\Delta^\pi(q, y) = (\lambda - \comp_\pi)q
        \label{eq:y-dept}
        &- \lambda \Delta_\pi(y) + \frac{1}{2} \lambda + \sum_{y', a} \mu^\pi_{y, y', a} \left(\frac{1}{2} (-a + u - \Delta_\pi(y'))^2 - \frac{1}{2} \Delta_\pi(y)^2 \right).
    \end{align}
\end{lemma}
\proof[Proof deferred to \ref{app:ak-lemmas}]{~}

We can now characterize the mean response time of the MMSR-$\pi$ system. 
We will use the fact that by \cref{lem:generator-f}, $G^\pi \circ f_\Delta^\pi(q, y)$ decomposes
into a term linearly dependent on the queue length $q$, and terms that are not dependent on $q$ except through the unused service $u$.
We define $c_0(y,q)$ to comprise the later group of terms.
We also define $c_1(y)$ and $c_2(y)$, which are simpler functions that are closely related to $c_0(y, q)$.
\begin{definition}
    \label{def:cs}
    Define $c_0(y, q), c_1(y),$ and $c_2(y)$ as follows:
    \begin{align*}
        c_0(y,q) &= G^\pi \circ f_\Delta^\pi(q, y) - (\lambda - \comppi)q \\
        &= - \lambda \Delta_\pi(y) + \frac{1}{2}\lambda  + \sum_{y', a} \mu^\pi_{y, y', a} \left(\frac{1}{2} (-a + u - \Delta_\pi(y'))^2 - \frac{1}{2} \Delta_\pi(y)^2 \right), \\
        c_1(y) &= - \lambda \Delta_\pi(y) + \frac{1}{2}\lambda  + \sum_{y', a} \mu^\pi_{y, y', a} \left(\frac{1}{2} (-a - \Delta_\pi(y'))^2 - \frac{1}{2} \Delta_\pi(y)^2 \right), \\
        c_2(y) &= c_1(y) - G^\pi \circ h(y), \text{ where }
        h(y) = \frac{1}{2} \Delta_\pi(y)^2 \\
        &= - \lambda \Delta_\pi(y) + \frac{1}{2}\lambda  + \sum_{y', a} \mu^\pi_{y, y', a} \left(\frac{1}{2} a^2 + a \Delta_\pi(y') \right).
    \end{align*}
\end{definition}
We will show that these functions' expected values,
$\Ep[c_0(Y^\pi, Q^\pi)], \Ep[c_1(Y^\pi)],$ and $\Ep[c_2(Y^\pi)]),$
are all equal up to a $O_\lambda(1-\frac{\lambda}{\comppi})$ error. This fact is crucial to our proof of \cref{thm:mmsr-response-time}.

\begin{reptheorem}{thm:mmsr-response-time}[Mean response time asymptotics of MMSR systems]
    In the MMSR-$\pi$ system, the expected response time in steady state satisfies
    \begin{equation}\label{eq:ak-response-time}
        \Ep[T^\pi] = \frac{1}{\comppi} \frac{1+\Delta_{\pi}(Y^\pi_d, Y^\pi)}{1-\lambda/\comppi}+O_\lambda(1).
    \end{equation}
\end{reptheorem}
\begin{proof}
In this proof we omit $\pi$ in the subscript of $\Delta_\pi(y)$ and in the superscript of $\mu^\pi_{y,y',a}$. 
We start from \cref{lem:generator-f}, which states that
\begin{align*}
    G^\pi \circ f(q, y) = (\lambda - \comp_\pi)q + c_0(y, q).
\end{align*}
Applying \cref{lem:drift-lemma}, we find that
\begin{align*}
    0&=\Ep[G^\pi \circ f(Q^\pi, Y^\pi)] = (\lambda - \comp_\pi)\Ep[Q^\pi] + \Ep[c_0(Y^\pi, Q^\pi)], \\
    \Ep[Q^\pi] &= \frac{\Ep[c_0(Y^\pi, Q^\pi)]}{\comppi - \lambda}.
\end{align*}
%
%
%
%
%
We therefore focus on $c_0(q,y)$: By characterizing $\Ep[c_0(Y^\pi, Q^\pi)]$,
we will characterize $\Ep[Q^\pi]$.

Let us separate out the terms where $u$ appears in $c_0(y, q)$ from the terms without $u$:
\begin{align}
    \label{eq:just-u}
    c_0(y, q) - c_1(y) = 
    \sum_{y',a} \mu_{y,y',a} u \left(
    \frac{1}{2}u
    -a
    -\Delta(y')
    \right).
\end{align}
Note that in the time-average steady state $Y^\pi$,
the fraction of service-process completions that occur while the queue is empty
(i.e. where $u=1$) is
$1-\frac{\lambda}{\comp_\pi}$,
because $\lambda$ jobs arrive per second,
and
$\comp$ service-process completions occur per second. As a result,
\begin{align*}
    E_{y \sim Y^\pi}\big[\sum_{y',a} \mu_{y, y', a} u\big] = 1-\frac{\lambda}{\comp_\pi}.
\end{align*}
Note that $a \le 1$ and $u \le 1$, because at most 1 job completes at a time.
Note that $\Delta(y')$ is bounded by a constant over all $y'$,
because $y' \in \yspi$, which is a finite state space.
Thus, the $u/2-a-\Delta(y')$ term in \eqref{eq:just-u} is bounded by a constant.
As a result, \eqref{eq:just-u}
contributes $O_\lambda(1-\frac{\lambda}{\comp_\pi})$ to $\Ep[c_0(Y^\pi, Q^\pi)]$:
\begin{align*}
    \nonumber
    &\hspace{-6pt} \Ep[c_1(Y^\pi) - c_0(Y^\pi, Q^\pi)] = O_\lambda(1-\lambda/\comp_\pi).
\end{align*}

Next, recall that $c_2(y) := c_1(y) - G^\pi \circ h(y)$.
By \cref{lem:drift-lemma}, $\Ep[G^\pi \circ h(Y^\pi)] = 0$,
so $\Ep[c_2(Y^\pi)] = \Ep[c_1(Y^\pi)]$.
Let us now simplify $c_2(y)$, using the fact that $a = 0$ or $1$:
\begin{align*}
    c_2(y) 
    &= - \lambda \Delta(y) + \frac{1}{2}\lambda  + \sum_{y', a} \mu^\pi_{y, y', a} \left(\frac{1}{2} a^2 + a \Delta_\pi(y') \right) \\
    &= -\lambda \Delta(y) + \frac{1}{2}\lambda +
    \frac{1}{2}\mu_{y,\cdot,1} +
    \sum_{y'} \mu_{y, y', 1}
    \Delta(y'). \nonumber
\end{align*}
We now apply \cref{lem:yd-distribution} to simplify the summation term of $c_2(y)$. \cref{lem:yd-distribution} states that
\begin{align*}
    \frac{1}{\comppi} \Ep_{y\sim Y^\pi}[\mu^\pi_{y,y',1}] = \Prob(Y^\pi_d = y').
\end{align*}
Thus, taking the expectation of the summation term of $c_2(y)$ over $y \sim Y^\pi$,
we find that
\begin{align*}
    \Ep_{y\sim Y^\pi}[\sum_{y'} \mu_{y, y', 1} \Delta(y')]
    &= \comp_\pi \sum_{y'} \Prob(Y^\pi_d = y') \Delta(y')
    = \comp_\pi \Delta(Y^\pi_d), \\
    \Ep[c_2(Y^\pi)] \hspace{20pt}&\hspace{-20pt}= \Ep[-\lambda \Delta(Y^\pi) + \frac{1}{2}(\mu_{Y^\pi, \cdot,1} + \lambda) + \comp_\pi \Delta(Y^\pi_d)].
\end{align*}
Now note that $\Ep[\Delta(Y^\pi)] = 0$, $\Ep[\mu_{Y^\pi,\cdot,1}] = \comp_\pi$, and $\lambda = \comp_\pi + O_\lambda(1-\frac{\lambda}{\comp_\pi})$:
\begin{align}
    \label{eq:c2-evaluated}
    \Ep[c_1(Y^\pi)] &= \Ep[c_2(Y^\pi)] = \comp_\pi + \comp_\pi \Delta(Y^\pi_d)
    +O_\lambda(1-\frac{\lambda}{\comp_\pi}).\\
    \nonumber
    \Ep[c_0(Y^\pi,Q^\pi)] &= \Ep[c_1(Y^\pi)] + O_\lambda(1-\frac{\lambda}{\comp_\pi})= \Ep[c_2(Y^\pi)]  
    +O_\lambda(1-\frac{\lambda}{\comp_\pi}). \\
    \nonumber
    \Ep[Q^\pi] &= \frac{\Ep[c_0(Y^\pi,Q^\pi)]}{\comppi-\comp}
    =\frac{\comp_\pi + \comp_\pi \Delta(Y^\pi_d)}{\comp_\pi - \lambda} + O_\lambda(1) =  
    \frac{\Delta(Y^\pi_d) + 1 }{1 - \lambda/\comp_\pi} + O_\lambda(1).
\end{align}
Now, we apply Little's Law, which states that $\Ep[T^\pi] = \frac{1}{\lambda} \Ep[Q^\pi]$:
\begin{align*}
    \Ep[T^\pi] &= \frac{1}{\lambda} \frac{1+\Delta(Y^\pi_d)}{1-\lambda/\comppi} + O_\lambda \left(\frac{1}{\lambda} \right).
\end{align*}
Note that for any $x$, $\frac{1}{\lambda} \frac{x}{1-\lambda/\comp} = \frac{1}{\comp} \frac{x}{1-\lambda/\comp} + \frac{x}{\lambda}$, so
\begin{align}
    \label{eq:mmsr-last}
    \Ep[T^\pi] = \frac{1}{\comp_\pi} \frac{1+\Delta(Y^\sat_d)}{1-\lambda/\comppi} + O_\lambda \left(\frac{1}{\lambda} \right). 
\end{align}

Note that in the $\lambda \to \comppi$ limit,
$O_\lambda(\frac{1}{\lambda}) = O_\lambda(1)$.
Consider the $\lambda \to 0$ limit: $\Ep[T^\pi]$ is bounded for small $\lambda$.
Likewise, $\frac{1}{\comppi} \frac{1+\Delta(Y^\pi_d)}{1-\lambda/\comppi}$ is bounded for small $\lambda$. As a result, the two differ by $O_\lambda(1)$:
\begin{align*}
    \Ep[T^\msj] = \frac{1}{\comppi} \frac{1+\Delta(Y^\pi_d)}{1-\lambda/\comppi} + O_\lambda(1).
    \qquad
    \qedhere
\end{align*}
\end{proof}




\section{RESET Proofs}
\label{sec:reset}

To characterize the asymptotic behavior of mean response time of the MSJ system,
we use the At-least-$k$ (Ak) system, which is stochastically equal to the MMSR-$\sat$ system.
The MARC results from \cref{sec:marc} allow us to characterize the MMSR-$\sat$ system.
To prove that the MSJ FCFS and Ak systems have the same asymptotic mean response time behavior,
our key idea is to show that $Y^\msj$ and $Y^\ak$,
the steady states of their fronts,
are ``almost identical."

To formalize and prove the relationship between $Y^\msj$ and $Y^\ak$,
we design a coupling in \cref{sec:coupling} between the MSJ system and the Ak system.
We use a renewal-reward argument based on busy periods to prove \cref{lem:tight-coupling}, which states
that under the coupling, $\Prob(Y^\msj \neq Y^\ak) = O_\lambda(1-\frac{\lambda}{\lambda^*})$.

Then, in \cref{sec:main-proof}, we combine \cref{thm:mmsr-response-time} and \cref{lem:tight-coupling} to prove \cref{thm:msj-response-time}, our main result, in which we give the first analysis of the asymptotic mean response time in the MSJ system,
by reduction to the saturated system.
\cref{thm:msj-response-time} parallels the proof steps that \cref{thm:mmsr-response-time}
uses to characterize the MMSR system,
using \cref{lem:tight-coupling}
to prove that the equivalent proof steps hold for the MSJ system.


We will make use of a test function $f^\msj_\Delta(q,y)$ for the multiserver-job system
which is similar to $f_\Delta^\pi(q,y)$, which was defined in \cref{def:f-delta}.
\begin{definition}
    \label{def:f-delta-msj}
    For states $y \in \ysak$, 
    \[
        f^\msj_\Delta(q, y) := f_\Delta^\ak(q, y)= f_\Delta^\sat(q, y).
    \]
    Otherwise,
    \[
        f^\msj_\Delta(q, y) := 0.
    \]
\end{definition}

Importantly, $G^\msj \circ f^\msj_\Delta(q, y)$ is similar to $G^\ak \circ f_\Delta^\ak(q, y)$:
\begin{lemma}
    \label{lem:generator-f-msj}
    \[
        G^\msj \circ f^\msj_\Delta(q, y) = \mathbbm{1}_{q > 0} G^\ak \circ f_\Delta^\ak(q, y) + \mathbbm{1}_{q=0} O_\lambda(1).
    \]
\end{lemma}
\proof[Proof deferred to \ref{app:generator-msj-lemmas}]{~}

\input{coupling-lemmas-proofs}

\subsection{Proof of \cref{thm:msj-response-time}}
\label{sec:main-proof}

We now are ready to prove our main theorem, \cref{thm:msj-response-time},
progressing along similar lines as \cref{thm:mmsr-response-time} and making use of \cref{lem:generator-f,lem:tight-coupling}.
%
%
First, we restate several definitions from \cref{def:cs}, specialized to $\ak$ system:
\begin{definition}\label{def:cs-ak}
    Recall the definitions of
    $c_0(y,q)$ and $c_1(y)$ 
    from \cref{def:cs}:
    \begin{align*}
        c_0(y,q) &= G^\ak \circ f_\Delta^\ak(q, y) - (\lambda - \comp)q \\
        &= - \lambda \Delta(y) + \frac{1}{2}\lambda  + \sum_{y', a} \mu_{y, y', a} \left(\frac{1}{2} (-a + u - \Delta(y'))^2 - \frac{1}{2} \Delta(y)^2 \right), \\
        c_1(y) &= - \lambda \Delta(y) + \frac{1}{2}\lambda  + \sum_{y', a} \mu_{y, y', a} \left(\frac{1}{2} (-a - \Delta(y'))^2 - \frac{1}{2} \Delta(y)^2 \right), 
    \end{align*}
    where $u=\mathbbm{1}\{q =0 \land a=1\}$. 
\end{definition}
We also make use of a key fact about $c_1(y)$, from \eqref{eq:c2-evaluated}:
\begin{align*}
    \Ep[c_1(Y^\ak)] &= \comp + \comp \Delta(Y^\sat_d)
    +O_\lambda \left(1-\frac{\lambda}{\comp} \right).
\end{align*}

Throughout this section, whenever we make use of results from \cref{sec:marc},
we set $\pi = \sat$.
In particular, we make use of $c_0(y, q)$ and $c_1(y)$, from \cref{def:cs}.

\begin{reptheorem}{thm:msj-response-time}
    In the multiserver-job system, the expected response time in steady state satisfies
    \begin{align*}
        \Ep[T^\msj] = \frac{1}{\lambda^*} \frac{1+\Delta(Y^\sat_d, Y^\sat)}{1-\lambda/\lambda^*}+O_\lambda(1).
    \end{align*}
\end{reptheorem}
\begin{proof}
    We will show that
the MSJ model has the same asymptotic mean response time as the Ak system.
We will make use of the test function $f_\Delta^\msj(q,y)$,
from \cref{def:f-delta-msj}.
Recall from \cref{lem:generator-f-msj} that
\begin{align*}
    G^\msj \circ f^\msj_\Delta(q, y) = G^\ak \circ f_\Delta^\ak(q, y) \mathbbm{1}_{q > 0} + \mathbbm{1}_{q=0} O_\lambda(1).
\end{align*}
We will next use \eqref{eq:generator-steady-zero}, the fact that the expected value of a generator function in steady state is zero, which implies that
\begin{align}
    \label{eq:msj-generator}
    0 &= \Ep[G^\ak \circ f_\Delta^\ak(Q^\msj, Y^\msj) \mathbbm{1}\{Q^\msj > 0\}] + \Prob(Q^\msj = 0) O_\lambda(1).
\end{align}
By \cref{lem:tight-coupling}, $\Prob(Q^\msj = 0) = O_\lambda(1-\frac{\lambda}{\lambda^*})$.
%
Next, we apply \cref{lem:generator-f} to the Ak system, finding that
\begin{align*}
    G^\ak \circ f_\Delta^\ak(q, y) = (\lambda - \comp)q + c_0(y, q).
\end{align*}

From \cref{def:cs-ak}, we can see that $c_0(y, q) \mathbbm{1}_{q>0} = c_1(y) \mathbbm{1}_{q>0}$.
Combining with \eqref{eq:msj-generator} and invoking \cref{lem:generator-f,lem:tight-coupling} and the fact that $c_1(y)$ is bounded, we have
\begin{align}
    \nonumber
    (\lambda - \comp) &\Ep[Q^\msj] + \Ep[c_0(Y^\msj, Q^\msj)\mathbbm{1}\{Q^\msj > 0 \}] = O_\lambda(1-\lambda/\lambda^*), \\
    \nonumber
    \nonumber
    (\lambda - \comp) &\Ep[Q^\msj] + \Ep[c_1(Y^\msj)] = O_\lambda(1-\lambda/\lambda^*), \\
    \label{eq:msj-first-q}
    &\Ep[Q^\msj] = \frac{\Ep[c_1(Y^\msj])}{\comp-\lambda} + O_\lambda(1).
\end{align}

Next, specializing \eqref{eq:c2-evaluated} in the proof of Theorem~\ref{thm:mmsr-response-time} to the Ak system,
we know that 
\begin{align*}\Ep[c_1(Y^\ak)] = \comp + \comp \Delta(Y^\sat_d)
    +O_\lambda \left(1-\frac{\lambda}{\comp} \right).
\end{align*}
By \cref{lem:tight-coupling}, we know that $\Prob(Y^\ak \neq Y^\msj) = O_\lambda(1-\frac{\lambda}{\lambda^*})$.
Again because $c_1(y)$ is bounded, 
\begin{align*}
    \Ep[c_1(Y^\msj)] = \Ep[c_1(Y^\ak)] + O_\lambda \left(1-\frac{\lambda}{\comp} \right) = \comp + \comp \Delta(Y^\sat_d)
    +O_\lambda \left(1-\frac{\lambda}{\comp} \right).
\end{align*}
Therefore, applying \eqref{eq:msj-first-q}, we find that 
\[
    \Ep[Q^\msj] = \frac{1 + \Delta(Y^\sat_d)}{1-\lambda/\lambda^*}
    + O_\lambda(1).
\]

Now, we apply Little's Law, which states that $\Ep[T^\ak] = \frac{1}{\lambda} \Ep[N^\ak]$.
Note that $Q^\ak$ and $N^\ak$ differ by the number of jobs in the front,  which is $O_\lambda(1)$:
\begin{align*}
    \Ep[T^\msj] &= \frac{1}{\lambda} \frac{1+\Delta(Y^\sat_d)}{1-\lambda/\lambda^*} + O_\lambda \left(\frac{1}{\lambda} \right) 
    = \frac{1}{\lambda^*} \frac{1+\Delta(Y^\sat_d)}{1-\lambda/\lambda^*} + O_\lambda \left( \frac{1}{\lambda} \right).
\end{align*}
For the second equality, note that for any $x$, $\frac{1}{\lambda} \frac{x}{1-\lambda/\comp} = \frac{1}{\comp} \frac{x}{1-\lambda/\comp} + \frac{x}{\lambda}$. Here $x$ is a constant, so the extra term is absorbed by the $O_\lambda(1/\lambda)$.

By the same bounding argument as used for \eqref{eq:mmsr-last} in the $\lambda \to 0$ limit,
\begin{align*}
    \Ep[T^\msj] = \frac{1}{\lambda^*} \frac{1+\Delta(Y^\sat_d)}{1-\lambda/\lambda^*} + O_\lambda(1).
    \qquad
    \qedhere
\end{align*}
\end{proof}




\section{Extensions of RESET: Finite skip models}
\label{sec:beyond}

While our main MSJ result, \cref{thm:msj-response-time}, was stated for the MSJ FCFS model,
our techniques do not depend on the details of that model.
Our RESET technique can handle a wide variety of models, which we call ``finite skip'' models:
\begin{definition}
    \label{def:finite-skip}
    A \emph{finite skip} queueing model is one in which jobs are served in near-FCFS order. Only jobs among the $n$ oldest jobs in arrival order are eligible for service, for some constant $n$.
    Service is only dependent on the states of the $n$ oldest jobs in arrival order,
    plus an optional environmental state from a finite-state Markov chain.
    Furthermore, jobs must have finite state spaces,
    and arrivals must be Poisson with i.i.d. initial job states.
\end{definition}
\cref{def:finite-skip} generalizes the work-conserving finite-skip (WCFS) class \cite{grosof_wcfs_2021}.
The MARC and RESET techniques can characterize the asymptotic mean response time
of \emph{any} finite skip model,
via the procedure in \cref{fig:structure}.
Additional finite skip MSJ models include nontrivial scheduling policies, including some backfilling policies; changing server need during service; multidimensional resource constraints; heterogeneous servers; turning off idle servers; and preemption overheads.
For discussion of each of these variants, see \ref{app:additional-msj}.

\section{Empirical Validation}
\label{sec:empirical}

We have characterized the asymptotic mean response time behavior of the FCFS multiserver-job system.
To illustrate and empirically validate our theoretical results,
we simulate the mean response time of the MSJ model
to compare it to our predictions.
Recall \eqref{eq:msj-response-time} from \cref{thm:msj-response-time},
in which we proved mean response time can be characterized as a dominant term plus a $O_\lambda(1)$ term:
\begin{align}
    \label{eq:msj-dominant}
    \Ep[T^\msj] = \frac{1}{\lambda^*} \frac{1+\Delta(Y^\sat_d, Y^\sat)}{1-\lambda/\lambda^*}+O_\lambda(1).
\end{align}
In this section, we simulate mean response time $\Ep[T^\msj]$,
and compare it against the dominant term of \eqref{eq:msj-dominant},
which we compute explicitly.

\subsection{Accuracy of formula}

\begin{figure}
    \centering
    \begin{subfigure}[b]{0.49\textwidth}
        \centering
        \includegraphics[width=\textwidth]{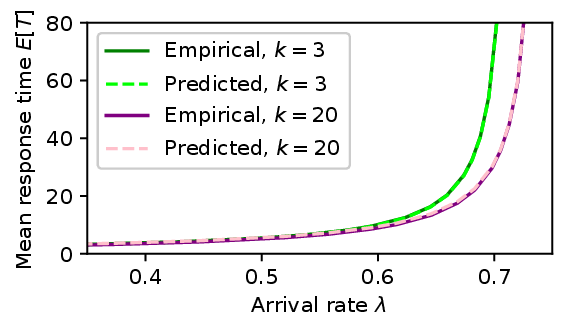}
        \caption{(1) $k=3$, server need sampled uniformly from $\{1, 2, 3\}$,
    durations $Exp(1/3), Exp(2/3),$ and $Exp(1),$ respectively.
    (2) $k=20$, server need sampled uniformly from $\{1, 20\}$,
    durations $Exp(1)$ and $Exp(1/2),$ respectively.}
        \label{fig:general_response}       
    \end{subfigure}
    \hfill
    \begin{subfigure}[b]{0.49\textwidth}
        \centering
        \includegraphics[width=\textwidth]{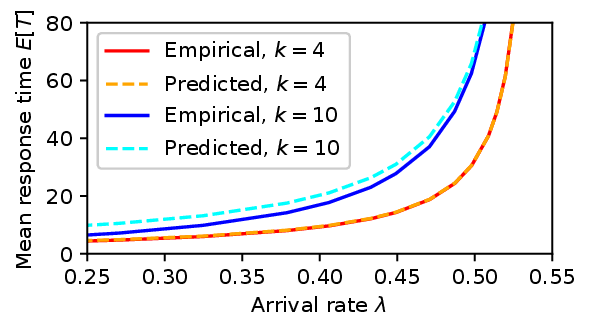}
        \caption{(1) $k=4$, two classes of jobs: Server need $1$, duration $Exp(1/4)$ w.p. $42\%$. Server need 4, duration $Exp(1)$ w.p. $58\%$. (2) $k=10$, two classes of jobs: Server need $1$, duration $Exp(1/10)$ w.p. $10\%$. Server need 10, duration $Exp(1)$ w.p. $90\%$.}
        \label{fig:matched_response}
    \end{subfigure}
    \caption{Empirical and predicted mean response time $\Ep[T]$
    for two MSJ settings in each of figures (a) and (b). Simulated $10^8$ arrivals at arrival rates ranging over $\lambda/\lambda^* \in [0.5, 0.99]$.}
    \label{fig:absolute-response}
\end{figure}

In \cref{fig:general_response}, we show that our predictions are an excellent match for the empirical behavior of the MSJ system in two different settings.
In the first, there are $k=3$ servers and jobs have server needs of 1, 2, and 3.
In the second, there are $k=20$ servers, and jobs have server needs 1 and 20. We thereby cover a spectrum from few-server-systems to many-server-systems, demonstrating extremely high accuracy in both regimes.
The $O_\lambda(1)$ term in \eqref{eq:msj-dominant} is negligible in both of these examples.

In \cref{fig:matched_response},
we compare mean response time in two settings with the same size distribution and stability region,
but which have very different $\Delta$.
We discuss these settings further in \cref{sec:empirical-understanding}.

The first setting has $k=4$, and $42\%$ of jobs have server need 1,
while $58\%$ of jobs have server need 4.
The second setting has $k=10$, and $10\%$ of jobs have server need 1,
while $90\%$ of jobs have server need 10.
The settings' stability regions are near-identical, with thresholds $\lambda^*_4 \approx 0.5413, \lambda^*_{10} \approx 0.5411$,
and their \emph{size} distributions, defined as duration times server need over $k$, are both $Exp(1)$.
However, our predictions for mean response time are very different in the two settings: $\Delta(Y_d^\sat)_4 \approx 0.3271, \Delta(Y_d^\sat)_{10} \approx 1.850$.
The $k=10$ setting considered here, with its relatively large value of $\Delta(Y_d^\sat)$,
is an especially difficult test-case.
Nonetheless, 
our predictions are validated by the simulation results in \cref{fig:matched_response}.

\begin{figure}
    \centering
    \includegraphics[width=\textwidth]{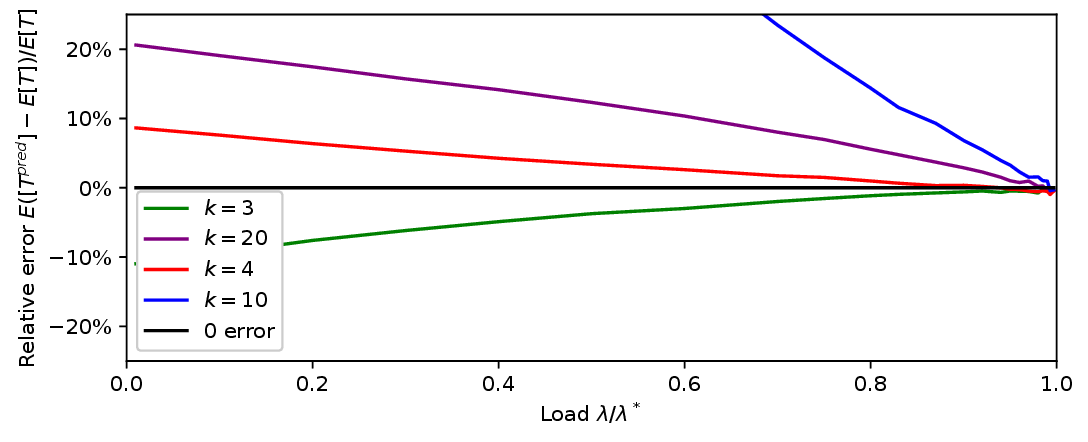}
    \caption{Relative error between empirical and predicted mean response time $E[T]$ for the four MSJ setting described in \cref{fig:absolute-response}. Simulated $10^8$ arrivals at arrival rates ranging over $\lambda/\comp \in [0,0.997]$.}
    \label{fig:relative-response}
\end{figure}

In \cref{fig:relative-response},
we illustrate the relative error between our predicted mean response time and the simulated mean response time for the four settings depicted in \cref{fig:absolute-response}.
In all four settings, as the arrival rate $\lambda$ approaches $\comp$, the threshold of the stability region, the relative error converges to 0.

Note that the convergence rate is slowest in the $k=10$ setting, which also has the largest $\Delta(Y_d^\sat)$ value.
We further explore the relationship between $\Delta(Y_d^\sat)$ values and convergence rates in \ref{app:delta-convergence}.
We find that such a correlation exists in some settings, but it is not robust or reliable.
\subsection{Understanding the importance of $\Delta$}
\label{sec:empirical-understanding}

Our results show that the relative completions function $\Delta$ is key to understanding the response time behavior of non-work-conserving systems such as the MSJ FCFS system.
This is in contrast to work-conserving systems, in which response time is determined
by the size distribution and load \cite{grosof_wcfs_2021}.
This contrast is illustrated by \cref{fig:matched_response},
in which we compare mean response time in two settings with the same size distribution and stability region,
but which have very different $\Delta$.

The differing mean response time behavior in these two settings
is caused by the difference in \emph{waste correlation}.
In the $k=10$ case, wasteful states persist for long periods of time:
If a 1-server job is the only job in service,
it takes more time for it to complete than in the $k=4$ system.
Thus, in the $k=4$ case, wasteful states are more short-lasting.
This difference in waste correlation produces the differences in $\Delta(Y_d^\sat)$
and in mean response time.

This example highlights a crucial feature of MSJ FCFS:
The failure of work conservation injects idiosyncratic idleness patterns in to the system.
To characterize $\Ep[T]$, we need to characterize these patterns, which the RESET and MARC techniques enable us to do for the first time.

\section{Conclusion}

We introduce the RESET and MARC techniques.
The RESET technique allows us to reduce the problem of characterizing mean response time in the MSJ FCFS system,
up to an additive constant,
to the problem of characterizing the M/M/1 with Markovian service rate (MMSR), where the service process is controlled by the saturated system.
The MARC technique gives the first explicit characterization of mean response time in the MMSR, up to an additive constant.
Together, our techniques reduce $\Ep[T^\msj]$ to two properties of the saturated system:
the departure-average steady state $Y_d^\sat$, and the relative completions function $\Delta(y_1, y_2)$.
Our RESET and MARC techniques apply to any finite skip model, including many MSJ generalizations.

We also introduce the simplified saturated system, a yet-simper variant of the saturated system with identical behavior.
We empirically validate our theoretical result,
showing that it closely tracks simulation at all arrival rates $\lambda$.


An important direction for future work is to analytically characterize the relative completions
$\Delta(y_1, y_2)$ for specific MSJ FCFS settings, such as settings where $Y_d^\sat$ is known to have a product-form distribution \cite{rumyantsev_2017,grosof_stability_2020}.

\section{Acknowledgements}
Isaac Grosof and Mor Harchol-Balter were supported by the National Science Foundation under grant number CMMI-2307008. Yige Hong was supported by the National Science Foundation under grant number ECCS-2145713. We thank the shepherd and the anonymous reviewers for their helpful comments.

\bibliographystyle{ACM-Reference-Format}
\bibliography{refs}

\appendix
\renewcommand{\thelemma}{\Alph{section}.\arabic{lemma}}
\renewcommand{\thecorollary}{\Alph{section}.\arabic{corollary}}

\input{appendix}

\section{Simplified Saturated System}
\label{app:sss}

For clarity, we refer to the previously-defined saturated system, defined in \cref{sec:saturated}, as the ``main saturated system.''

While the main saturated system is a finite-state system, it can have a very large number of possible states.
We therefore introduce the \emph{simplified saturated system} (SSS),
a new closed system with identical behavior, but smaller state space.
The SSS can be more amenable to theoretical analysis,
such as in the case of the product-form result in \cite{grosof_stability_2020}.

The simplified saturated system is a closed system which always contains jobs with total server need $\ge k$, and contains the minimal number of jobs to reach that threshold.
Whenever a job completes, the system admits new jobs until the total server need is $\ge k$. Jobs are served in FCFS order. Note that at most one job in the system is not in service.

In particular, a state of the SSS consists of a multiset of job states for the jobs in service,
plus the server need of the job not in service, if any.
The total server need of these jobs is just enough to be $\ge k$.

For instance, consider a system with $k=30$ jobs, and server needs either $3$ or $10$,
and exponential durations.
The main saturated system has state space $\y^\sat = \{3,10\}^{30}$, with over a billion states.
In contrast, the simplified saturated system has 13 states. We will write each state as a triple,
consisting of the server need of the job not in service, and the number of $3$-server and $10$-server jobs in service. Then the state space of the SSS is:
\begin{align*}
    \y^\sss = \{ &[\emptyset, 0, 3], [10, 1, 2], [10, 2, 2], [3, 3, 2], [10, 3, 2], [10, 4, 1], [10, 5, 1], \\ 
    &[3, 6, 1], [10, 6, 1], [10, 7, 0], [10, 8, 0], [10, 9, 0], [\emptyset, 10, 0]\}.
\end{align*}

Despite its much smaller state space, the SSS has essentially identical behavior to the main saturated system:

\begin{replemma}{lem:sss-coupling}
    There exists a coupling under which the main saturated system
    and simplified saturated system
    have identical completions.
\end{replemma}
\begin{proof}
    To form the coupling, let us sample in advance the entire arrival sequence: For each arrival, we pre-sample which initial state it will arrive in.

    Next, we initialize both systems based on this arrival sequence: For the main saturated system,
    the first $k$ jobs are initially present, while for the simplified saturated system,
    a subset of those jobs are initially present.
    Note that the set of jobs in service in the main saturated system is identical to the set of jobs in service in the simplified saturated system,
    because the total server need of jobs in service is at most $k$.
    Note that the ordering of the jobs in service does not affect any transitions,
    so the fact that SSS does not track this information poses no obstacle.
    We will ensure that the set of jobs in service in the two systems is identical throughout time.

    Next, we couple the two systems' completions and job state transitions to be identical.
    Jobs' states can only change while those jobs are in service,
    so this coupling is valid as long as the set of jobs in service is identical in both systems.
    Finally, whenever a pair of jobs completes, new jobs are generated according to the shared global arrival sequence. This ensures that the jobs that enter service are identical in the two systems.

    By construction, the set of jobs in service is always identical in the two systems.
    Under this coupling, the completion moments are also identical in the two systems.
\end{proof}


\section{Closed-form formulas for $\lambda^*, Y_d, \Delta(y)$}
\label{app:calculate}

Our result on the M/M/1 with Markovian Service Rate (MMSR), \cref{thm:mmsr-response-time},
characterizes mean response time in the MMSR-$\pi$ system in terms of the following quantities:
\begin{itemize}
    \item $\comppi$, the threshold of the stability region of the MMSR-$\pi$ system,
    \item $Y_d^\pi$, the departure-average steady state of the service process $\pi$, and
    \item $\Delta_\pi(y)$, the relative completions function of the service process $\pi$.
\end{itemize}

Similarly, our result on the MSJ system, \cref{thm:msj-response-time},
characterizes mean response time in the MSJ system in terms of $\comp, Y_d^\sat,$ and $\Delta_\sat(y)$,
or equivalently in terms of $\comp, Y_d^\sss,$ and $\Delta_\sss(y)$, by \cref{cor:msj-sss-response-time}.

In this section, we demonstrate how to explicitly calculate $\comppi, Y_d^\pi$, and $\Delta_\pi(y)$
by solving a system of linear equations,
and walk through this exercise for a specific parameterized setting,
giving an explicit, closed-form expression for mean response time within the given parameterized setting.

\subsection{Solving for $\comppi, Y_d^\pi,$ and $\Delta_\pi(y)$}

First, we solve the continuous-time balance equations for the service process $\pi$ to determine
the time-average steady state $Y^\pi$:
\begin{align}
    \label{eq:calc-time-steady}
    \forall y \in \yspi, \quad \Prob(Y^\pi = y) \mu_{y,\cdot,\cdot} &= \sum_{y' \in \yspi} \Prob(Y^\pi = y') \mu_{y',y,\cdot}, \\
    \nonumber
    \sum_{y \in \yspi} \Prob(Y^\pi = y) &= 1.
\end{align}

Next, we calculate the throughput $X^\pi$ of the service process $\pi$,
which by prior results \cite{baccelli_1995} is equal to the threshold of the MMSR-$\pi$ stability region,
$\comppi$:
\begin{align}
    \label{eq:calc-comppi}
    \comppi = X^\pi = \Ep_{y \sim Y^\pi}[\mu_{y,\cdot,1}] = \sum_{y \in \ysak} \mu_{y,\cdot,1} \Prob(Y^\pi = y).
\end{align}

Next, we calculate the departure-average steady state $Y_d^\pi$.
Recall that $Y_d^\pi$ is the steady-state distribution of the embedded DTMC
which samples states just after each departure from $\pi$.
To calculate $\Prob(Y^\pi_d = y)$ from $\Prob(Y^\pi = d)$,
we divide by the expected time spent in state $y$ per visit, $\frac{1}{\mu_{y,\cdot,\cdot}}$,
and multiply by the probability that the transition into state $y$ was a completion:
\begin{align}
    \label{eq:calc-dept}
    \Prob(Y^\pi_d = y) = Z^\pi \Prob(Y^\pi = y) \frac{\mu_{\cdot,y,1}}{\mu_{\cdot,y,\cdot}} \mu_{y,\cdot,\cdot},
\end{align}
where $Z^\pi$ is a normalization constant.

Finally, we calculate the relative completions function $\Delta_\pi(y)$.
To do so, we use the system of equations given in \cref{cor:forward-delta}:
\begin{align}
    \label{eq:calc-delta-main}
    \Delta_\pi(y) = \frac{\mu_{y,\cdot,1} - \comppi}{\mu_{y,\cdot,\cdot}} + \sum_{y'}\frac{\mu_{y,y',\cdot}}{\mu_{y,\cdot,\cdot}} \Delta_\pi(y').
\end{align}

This system of equations characterizes $\Delta_\pi(y)$ up to an additive offset. To uniquely determine $\Delta_\pi(y)$,
we use the fact that $\Delta(Y^\pi) = 0$:
\begin{align}
    \label{eq:calc-delta-sum}
    0 = \Delta(Y^\pi) = \sum_y \Prob(Y^\pi = y) \Delta(y).
\end{align}

\subsection{Specific parameterized example: Two servers, arbitrary service rates}
\label{app:calculate-k2}

We now walk through the process of determining $\comp, Y_d^\sss,$ and $\Delta_\sss(y)$,
in a parameterized MSJ setting, allowing us to use \cref{thm:msj-response-time} to explicitly characterize mean response time $\Ep[T^\msj]$.

Consider an MSJ system with $k=2$ servers, and jobs with server need either 1 or 2.
$p_1$ fraction of jobs have server need 1,
and $p_2 = 1-p_1$ have server need 2.
Let server need 1 jobs have service duration $Exp(\mu_1)$, and server need 2 jobs have service duration $Exp(\mu_2)$.

Note that this setting is a generalized, parameterized version of the setting discussed in \cref{sec:empirical}.
The same methods can handle any MSJ setting with phase-type service durations - we merely choose this one as a clean example.

The Simplified Saturated System (SSS) for this setting has three states: $[1,1]$, $[1, 2]$, and $[2]$.
Between these states, we have the following transition rates:
\begin{align*}
    \mu_{[1,1],[1,1],1} &= 2 \mu_1 p_1, \quad \mu_{[1,1],[1,2],1} = 2 \mu_1 p_2, \quad \mu_{[1,2],[2],1} = \mu_1, \\
    \mu_{[2],[1,1],1} &= \mu_2 p_1^2, \quad \mu_{[2],[1,2],1} = \mu_2 p_1 p_2, \quad \mu_{[2],[2],1]} = \mu_2 p_2.
\end{align*}
Note that all transitions in this setting involve a completion.
This is due to the fact that the service durations are exponential.
For more complex service duration distributions, there would also be non-completion transitions.

Now, we calculate the time-average steady state $Y$, using \eqref{eq:calc-time-steady}:
\begin{align*}
    2 \mu_1 \Prob(Y = [1,1]) &= 2\mu_1 p_1 \Prob(Y = [1,1]) + \mu_2 p_1^2 \Prob(Y = [2]), \\
    \mu_1 \Prob(Y = [1,2]) &= 2\mu_1 p_2 \Prob(Y = [1,1]) + \mu_2 p_1 p_2 \Prob(Y = [2]), \\
    \mu_2 \Prob(Y = [2]) &= \mu_1 \Prob(Y=[1,2]) + \mu_2 p_2 \Prob(Y = [2]), \\
    &\Prob(Y = [1,1]) + \Prob(Y=[1,2]) + \Prob(Y=[2]) = 1.
\end{align*}

Solving, we find that
\begin{align*}
    \Prob(Y=[1,1]) &= \frac{\mu_2 p_1^2}{\mu_2 p_1^2 + 2 \mu_2 p_1 p_2 + 2 \mu_1 p_2}, \\
    \Prob(Y=[1,2]) &= \frac{2 \mu_2 p_1 p_2}{\mu_2 p_1^2 + 2 \mu_2 p_1 p_2 + 2 \mu_1 p_2}, \\
    \Prob(Y=[2]) &= \frac{2 \mu_1 p_2}{\mu_2 p_1^2 + 2 \mu_2 p_1 p_2 + 2 \mu_1 p_2}.
\end{align*}

Next, we use \eqref{eq:calc-comppi} to calculate $\comp$, the threshold of the stability region:
\begin{align*}
    \comp = 2 \mu_1 \Prob(Y=[1,1]) + \mu_1 \Prob(Y=[1,2]) + \mu_2 \Prob(Y=[2]) = \frac{2 \mu_1 \mu_2}{\mu_2 p_1^2 + 2 \mu_2 p_1 p_2 + 2 \mu_1 p_2}.
\end{align*}

Next, we use \eqref{eq:calc-dept} to calculate $Y_d$, the departure-average steady-state.
Note that all transitions are completions, so \eqref{eq:calc-dept} simplifies to the following:
\begin{align}
 \nonumber   \Prob(Y_d = y) &= \frac{1}{\comp} \Prob(Y = y) \mu_{y,\cdot,\cdot}, \\
 \label{eq:calc-yd}   \Prob(Y_d = [1,1]) &= p_1^2, \\
 \nonumber   \Prob(Y_d = [1,2]) &= p_1 p_2, \\
 \nonumber   \Prob(Y_d = [2]) &= p_2.
\end{align}
Note that this is a product-form distribution. This is a special case of the product-form behavior that was established for the general 2-class exponential MSJ setting \cite{grosof_new_2023,grosof_stability_2020}.

Finally, we use \eqref{eq:calc-delta-main} and \eqref{eq:calc-delta-sum} to derive $\Delta(y)$ for each state $y$.
Because all transitions are completions, \eqref{eq:calc-delta-main} simplifies to the following:
\begin{align*}
    \Delta(y) &= 1 - \frac{\comp}{\mu_{y,\cdot,\cdot}} + \sum_{y'} \frac{\mu_{y,y',\cdot}}{\mu_{y,\cdot,\cdot}} \Delta(y').
\end{align*}

Now, let's substitute in our expressions for $\comp$ and the transition rates and simplify. First we characterize $\Delta([1,1])$:
\begin{align}
 \nonumber
    \Delta([1,1]) &= 1 - \frac{1}{2 \mu_1} \frac{2 \mu_1 \mu_2}{\mu_2 p_1^2 + 2 \mu_2 p_1 p_2 + 2 \mu_1 p_2}
    + p_1 \Delta([1,1]) + p_2 \Delta([1,2]), \\ \nonumber
    p_2 \Delta([1,1]) &= 1 - \frac{\mu_2}{\mu_2 p_1^2 + 2 \mu_2 p_1 p_2 + 2 \mu_1 p_2} + p_2 \Delta([1,2]), \\ \nonumber
    \Delta([1,1]) &= \frac{1}{p_2} \frac{\mu_2 p_1^2 + 2 \mu_2 p_1 p_2 + 2 \mu_1 p_2 - \mu_2}{\mu_2 p_1^2 + 2 \mu_2 p_1 p_2 + 2 \mu_1 p_2} + \Delta([1,2]) \\ \nonumber
    &= \frac{1}{p_2} \frac{\mu_2 p_1 p_2 + (2 \mu_1 - \mu_2) p_2}{\mu_2 p_1^2 + 2 \mu_2 p_1 p_2 + 2 \mu_1 p_2} + \Delta([1,2]) \\  \nonumber
    &= \frac{\mu_2 p_1 + 2 \mu_1 - \mu_2}{\mu_2 p_1^2 + 2 \mu_2 p_1 p_2 + 2 \mu_1 p_2} + \Delta([1,2]) \\
    \label{eq:calc-delta11}
    &= \frac{2 \mu_1 - \mu_2 p_2}{\mu_2 p_1^2 + 2 \mu_2 p_1 p_2 + 2 \mu_1 p_2} + \Delta([1,2]). 
\end{align}
Next $\Delta([1,2])$:
\begin{align}
\nonumber
    \Delta([1,2]) &= 1 - \frac{1}{\mu_1} \frac{2 \mu_1 \mu_2}{\mu_2 p_1^2 + 2 \mu_2 p_1 p_2 + 2 \mu_1 p_2} + \Delta([2]) \\ \nonumber
    &= 1 - \frac{2 \mu_2}{\mu_2 p_1^2 + 2 \mu_2 p_1 p_2 + 2 \mu_1 p_2} + \Delta([2]) \\ \nonumber
    &= \frac{\mu_2 p_1^2 + 2 \mu_2 p_1 p_2 + 2 \mu_1 p_2 - 2 \mu_2}{\mu_2 p_1^2 + 2 \mu_2 p_1 p_2 + 2 \mu_1 p_2} + \Delta([2]) \\ \label{eq:calc-delta12}
    &= \frac{-\mu_2 p_1^2 + 2 (\mu_1 - \mu_2) p_2}{\mu_2 p_1^2 + 2 \mu_2 p_1 p_2 + 2 \mu_1 p_2} + \Delta([2])
\end{align}
Finally, we characterize $\Delta([2])$:
\begin{align}
    \nonumber
    \Delta([2]) &= 1 - \frac{1}{\mu_2} \frac{2 \mu_1 \mu_2}{\mu_2 p_1^2 + 2 \mu_2 p_1 p_2 + 2 \mu_1 p_2} + p_1^2 \Delta([1,1]) + p_1 p_2 \Delta([1,2]) + p_2 \Delta([2]), \\ \nonumber
    p_1 \Delta([2]) &= 1 - \frac{2 \mu_1}{\mu_2 p_1^2 + 2 \mu_2 p_1 p_2 + 2 \mu_1 p_2} + p_1^2 \Delta([1,1]) + p_1 p_2 \Delta([1,2]), \\ \nonumber
    \Delta([2]) &= \frac{1}{p_1}\frac{\mu_2 p_1^2 + 2 \mu_2 p_1 p_2 + 2 \mu_1 p_2 - 2 \mu_1}{\mu_2 p_1^2 + 2 \mu_2 p_1 p_2 + 2 \mu_1 p_2} + p_1 \Delta([1,1]) + p_2 \Delta([1,2]) \\ \nonumber
    &= \frac{1}{p_1}\frac{(\mu_2 - 2 \mu_1) p_1^2 + 2 (\mu_2 - \mu_1) p_1 p_2}{\mu_2 p_1^2 + 2 \mu_2 p_1 p_2 + 2 \mu_1 p_2} + p_1 \Delta([1,1]) + p_2 \Delta([1,2]) \\ \label{eq:calc-delta2}
    &= \frac{(\mu_2 - 2 \mu_1) p_1 + 2 (\mu_2 - \mu_1) p_2}{\mu_2 p_1^2 + 2 \mu_2 p_1 p_2 + 2 \mu_1 p_2} + p_1 \Delta([1,1]) + p_2 \Delta([1,2]).
\end{align}

Note that our final equations, \eqref{eq:calc-delta11}, \eqref{eq:calc-delta12}, and \eqref{eq:calc-delta2},
are redundant: We can omit any one and still calculate $\Delta(y)$, up to an additive constant.
From these equations, we find that
\begin{align*}
    \Delta([1,1]) &= \frac{2 \mu_1 - \mu_2 p_2}{\mu_2 p_1^2 + 2 \mu_2 p_1 p_2 + 2 \mu_1 p_2} + C, \\
    \Delta([1,2]) &= C, \\
    \Delta([2]) &= \frac{\mu_2 p_1^2 + 2 (\mu_2 - \mu_1) p_2}{\mu_2 p_1^2 + 2 \mu_2 p_1 p_2 + 2 \mu_1 p_2} + C,
\end{align*}
where $C$ is an additive constant to be determined. To find $C$, we use \eqref{eq:calc-delta-sum}.
Substituting our known values, we find that
\begin{align*}
    0 &= \frac{\mu_2 p_1^2 (2 \mu_1 - \mu_2 p_2)}{(\mu_2 p_1^2 + 2 \mu_2 p_1 p_2 + 2 \mu_1 p_2)^2}
    + \frac{2 \mu_1 p_2 (\mu_2 p_1^2 + 2 (\mu_2 - \mu_1) p_2)}{(\mu_2 p_1^2 + 2 \mu_2 p_1 p_2 + 2 \mu_1 p_2)^2} + C, \\
    -C (\mu_2 p_1^2 + 2 \mu_2 p_1 p_2 + 2 \mu_1 p_2)^2 &= \mu_2 p_1^2 (2 \mu_1 - \mu_2 p_2) + 2 \mu_1 p_2 (\mu_2 p_1^2 + 2 (\mu_2 - \mu_1) p_2) \\
    &= -4 \mu_1^2 p_2^2 + 2\mu_1 \mu_2 (p_1^2 + p_1^2 p_2 + 2 p_2^2) - \mu_2^2 p_1^2 p_2, \\
    C &= \frac{4 \mu_1^2 p_2^2 - 2\mu_1 \mu_2 (p_1^2 + p_1^2 p_2 + 2 p_2^2) + \mu_2^2 p_1^2 p_2}{(\mu_2 p_1^2 + 2 \mu_2 p_1 p_2 + 2 \mu_1 p_2)^2}.
\end{align*}
We can therefore derive expressions for $\Delta(y)$:
\begin{align*}
    \Delta([1,1]) &= \frac{2 p_2 (2 \mu_1^2 (1+p_2) - \mu_1 \mu_2(-2 p_1 + p_1^2 + 3 p_2) - \mu_2^2 p_1 p_2)}{(\mu_2 p_1^2 + 2 \mu_2 p_1 p_2 + 2 \mu_1 p_2)^2}, \\
    \Delta([1,2]) &= \frac{4 \mu_1^2 p_2^2 - 2\mu_1 \mu_2 (p_1^2 + p_1^2 p_2 + 2 p_2^2) + \mu_2^2 p_1^2 p_2}{(\mu_2 p_1^2 + 2 \mu_2 p_1 p_2 + 2 \mu_1 p_2)^2}, \\
    \Delta([2]) &= \frac{\mu_2 p_1 (-2 \mu_1(1+p_2^2) + \mu_2 (p_1^2 p_1^2 p_2 + 3 p_2 + p_2^2))}{(\mu_2 p_1^2 + 2 \mu_2 p_1 p_2 + 2 \mu_1 p_2)^2}.
\end{align*}

Finally, we can apply our expressions for $Y_d$, \eqref{eq:calc-yd}, to calculate $\Delta(Y_d)$:

\begin{align*}
    \Delta(Y_d) &= \Ep_{y \sim Y_d}[\Delta(y)] \\
    &= \Delta([1,1]) \Prob(Y_d = [1,1]) + \Delta([1,2]) \Prob(Y_d = [1,2]) + \Delta([2]) \Prob(Y_d = [2]) \\
    &= p_1^2 \Delta([1,1]) + p_1 p_2 \Delta([1,2]) + p_2 \Delta([2]) \\
    &= \frac{p_1 p_2 (4 \mu_1^2 - 2 \mu_1 \mu_2 (1+3 p_2)+\mu_2^2 (1+p_2+2p_2^2))}{(\mu_2 p_1^2 + 2 \mu_2 p_1 p_2 + 2 \mu_1 p_2)^2}.
\end{align*}

Having explicitly characterized $\comp$ and $\Delta(Y_d)$, our main result, \cref{thm:msj-response-time}, gives an explicit, closed-form expression for mean response time:
\begin{align*}
    \Ep[T^\msj] = \frac{1}{\lambda^*} \frac{1+\Delta(Y_d)}{1-\lambda/\lambda^*}+O_\lambda(1).
\end{align*}

\section{MMSR Lemmas}
\label{app:ak-lemmas}
\begin{lemma}\label{lem:generator-delta}
    Consider the MMSR system controlled by the Markov chain $\pi$. For any state $y \in \yspi$,
    \begin{equation}
        G^\pi \circ \Delta_{\pi}(y, Y^\pi) = \comp_\pi - \mu^\pi_{y,\cdot,1}.
    \end{equation}
\end{lemma}
\begin{proof}
    Recall that by the definition of the generator, $G^\pi \circ  \Delta_\pi(y, Y^\pi)$ is given by
    \begin{equation}\label{eq:def-gak-deltay}
        G^\pi \circ  \Delta_\pi(y, Y^\pi) = \lim_{t \to 0} \frac{1}{t} \Ep[\Delta_\pi(Y^\pi(t), Y^\pi) - \Delta_\pi(y, Y^\pi) | Y^\pi(0)=y].
    \end{equation}
    To figure out $\Ep[\Delta_\pi(Y^\pi(t), Y^\pi) - \Delta_\pi(y, Y^\pi) | Y^\pi(0)=y]$, recall the definition that
    \[
        \Delta_\pi(y,Y^\pi) = \lim_{t' \to \infty} \Ep[C_\pi(y, t') - \comp_\pi t'],
    \]
    where recall that $C_\pi(y, t')$ is the expected number of completion up to time $t'$ of the MMSR system whose service process is controlled by the Markov chain $\pi$ initializing in state $y$. 
    Therefore, if we replace $y$ by $Y^\pi(t)$ on the LHS of the above definition and take the expectation, we have 
    \begin{align*}
        &\mspace{20mu} \Ep[\Delta_\pi(Y^\pi(t), Y^\pi)| Y^\pi(0)=y] \\
        &=  \lim_{t' \to \infty} \Ep[C_\pi(Y^\pi(t), t') - \comp_\pi t' | Y^\pi(0)=y ] \\
        & = \lim_{t'\to\infty} \Ep[C_\pi(y, t+t') - C_\pi(y, t) - \comp_\pi t' | Y^\pi(0)=y ],
    \end{align*}
    where in the second equality we have used the fact that 
    \begin{align}
        \label{eq:split-completions}
        \Ep[C_\pi(y, t+t')] &= \Ep[C_\pi(y,t) + C_\pi([Y_\pi(t) \mid Y_\pi(0) = y], t')] \\
        \nonumber
         \Ep[C_\pi(Y_\pi(t), t') \mid Y_\pi(0) = y] &= \Ep[C_\pi(y, t+t')] - \Ep[C_\pi(y,t)].
    \end{align}
    \eqref{eq:split-completions} simply splits up the completions from time 0 to $t+t'$ into the completions from time 0 to $t$, and the completions from time $t$ to $t+t'$.

    Therefore,
    \begin{align*}
        &\mspace{20mu} \Ep[\Delta_\pi(Y^\pi(t), Y^\pi)) - \Delta_\pi(y, Y^\pi) | Y^\pi(0)=y] \\
        &= \lim_{t'\to\infty} \Ep[C_\pi(y, t+t') - C_\pi(y, t) - \comp_\pi t'] - \lim_{t'\to\infty} \Ep[C_\pi(y, t') - \comp_\pi t'] \\
        &= \lim_{t'\to\infty} \Ep[C_\pi(y, t+t') - C_\pi(y, t) - \comp_\pi t'] - \lim_{t'\to\infty} \Ep[C_\pi(y, t+t') - \comp_\pi t - \comp_\pi t'] \\
        &= \Ep[-C_\pi(y, t) + \comp_\pi t],
    \end{align*}
    where in the second inequality we replace $t'$ with $t+t'$ in the second term, which will not change the limit because $t'$ and $t+t'$ are both going to infinity. Plugging the above calculations into \eqref{eq:def-gak-deltay}, we get 
    \begin{align*}
        G^\pi \circ  \Delta_\pi(y, Y^\pi) &= \lim_{t \to 0} \frac{1}{t} \Ep[-C_\pi(y, t) + \comp_\pi t] = - \mu^\pi_{y, \cdot, 1} +  \comp_\pi,
    \end{align*}
    where in the last inequality we use the fact that $\lim_{t\to 0}\frac{1}{t} \Ep[C_\pi(y, t)] = \mu^\pi_{y, \cdot, 1}$ (the instantaneous completion rate at state $y$).
\end{proof}

\begin{lemma}\label{lem:generator-ak}
    For any $f(q,y)$ which is a real-valued function of the state of the MMSR-$\pi$ system,
    \begin{align*}
        G^\pi \circ f(q,y) =  \lambda \left(f(q+1, y) - f(q, y) \right) + \sum_{\substack{y'\in\yspi,\\ a\in\{0,1\}}} \mu^\pi_{y, y', a} \left(f((q-a)^+, y') - f(q, y)\right).
    \end{align*}
\end{lemma}
\begin{proof}
    In this proof we omit $\pi$ in the subscript of $\Delta_\pi(y)$ and in the superscript of $\mu^\pi_{y,y',a}$ for readability.
    
    Recall the definition of the generator
    \[
        G^\pi \circ f(q, y) = \lim_{t \to 0} \frac{1}{t} \Ep[f(Q^\pi(t), Y^\pi(t)) - f(q, y) | Q^\pi(0) = q, Y^\pi(0)=y],
    \]
    which can be interpreted as the instantaneous rate of change of the function $f(Q^\pi(t), Y^\pi(t))$ when $(Q^\pi(t), Y^\pi(t))$ is initialized in $(q, y)$. 
    Note that $(Q^\pi(t), Y^\pi(t))$ can change either due to an arrival event, or a transition event of the Markov chain $\pi$.
    An arrival event happens with rate $\lambda$, and causes $Q^\pi(t)$ to change from $q$ to $q+1$, so arrival events contribute
    \[
        \lambda \left(f(q+1, y) - f(q, y) \right)
    \]
    to $G^\pi \circ f(q,y)$.
    A transition event of the Markov chain $\pi$ from $y$ to $y' \in \yspi$ accompanied by $a \in \{0, 1\}$ completions happens with rate $\mu_{y, y', a}$. Such a event causes $(Q^\pi(t), Y^\pi(t))$ to change from $(q, y)$ to $((q-a)^+, y')$, so it contributes
    \[
        \mu_{y, y', a} \left(f((q-a)^+, y') - f(q, y)\right)
    \]
    to $G^\pi \circ f(q,y)$, for each $y'\in \ysak$ and $a\in \{0, 1\}$. This proves the expression in the lemma statement. 
\end{proof}

As a corollary of \cref{lem:generator-delta} and \cref{lem:generator-ak},
we can derive a forward recurrence for $\Delta_\pi(y) := \Delta_\pi(y,Y^\pi).$
Solving the resulting system of equations, together with the fact that $\Delta_\pi(Y^\pi)=0$,
gives the value of $\Delta_\pi(y)$.

\begin{corollary}\label{cor:forward-delta}
    For any MMSR-$\pi$ system and any state $y \in \yspi$,
    \begin{align*}
        \Delta_\pi(y) = \frac{\mu_{y,\cdot,1} - \comppi}{\mu_{y,\cdot,\cdot}}
        + \sum_{y'} \frac{\mu_{y,y',\cdot}}{\mu_{y,\cdot,\cdot}} \Delta(y'),
    \end{align*}
    where $\mu_{y,\cdot,\cdot}$ is the total transition rate out of state $y$.
    
    Moreover, if all transitions in $\pi$ are associated with completions
    (if $a$ always equals $1$),
    then the recurrence simplifies:
    \begin{align*}
        \Delta_\pi(y) = 1 - \frac{\comppi}{\mu_{y,\cdot,1}}
        + \sum_{y'} \frac{\mu_{y,y',1}}{\mu_{y,\cdot,1}} \Delta(y').
    \end{align*}
\end{corollary}
\begin{proof}
    Start with \cref{lem:generator-delta}:
    \begin{align}
        \label{eq:forward-start}
        G^\pi \circ \Delta_{\pi}(y) = \comp_\pi - \mu^\pi_{y,\cdot,1}.
    \end{align}
    Here we write $\Delta_\pi(y)$ as a shorthand for $\Delta_\pi(y,Y^\pi).$
    
    Expand the left-hand side of \eqref{eq:forward-start} using \cref{lem:generator-ak}:
    \begin{align*}
        G^\pi \circ \Delta_{\pi}(y) = \sum_{y', a} \mu^\pi_{y,y',a} (\Delta_{\pi}(y') - \Delta_{\pi}(y)).
    \end{align*}
    Note that \cref{lem:generator-ak} simplifies because $\Delta_{\pi}(y)$ does not depend on $q$.

    Now we can perform algebraic manipulation to complete the proof:
    \begin{align*}
        \comp_\pi - \mu^\pi_{y,\cdot,1} &= \sum_{y', a} \mu^\pi_{y,y',a} (\Delta_{\pi}(y') - \Delta_{\pi}(y)) \\
        &= -\mu^\pi_{y,\cdot,\cdot} \Delta_{\pi}(y) + \sum_{y', a} \mu^\pi_{y,y',a} \Delta_{\pi}(y'), \\
        \mu^\pi_{y,\cdot,\cdot} \Delta_{\pi}(y) &=
        \mu^\pi_{y,\cdot,1} - \comp_\pi  + \sum_{y', a} \mu^\pi_{y,y',a} \Delta_{\pi}(y'), \\
        \Delta_{\pi}(y) &=
        \frac{\mu^\pi_{y,\cdot,1} - \comp_\pi}{\mu^\pi_{y,\cdot,\cdot} }  + \sum_{y', a} \frac{\mu^\pi_{y,y',a}}{\mu^\pi_{y,\cdot,\cdot}} \Delta_{\pi}(y').
    \end{align*}
    Note that if all transitions are associated with completions, e.g. if $a=1$, then
    $\mu^\pi_{y,\cdot,1} = \mu^\pi_{y,\cdot,\cdot}$
\end{proof}

\begin{replemma}{lem:generator-f}
    For any state $(q, y)$ of the MMSR-$\pi$ system,
    \begin{align}
        G^\pi \circ f_\Delta^\pi(q, y) &= (\lambda - \comp_\pi)q  - \lambda \Delta_\pi(y) + \frac{1}{2} \lambda+ \sum_{y', a} \mu^\pi_{y, y', a} \left(\frac{1}{2} (-a + u - \Delta_\pi(y'))^2 - \frac{1}{2} \Delta_\pi(y)^2 \right).
    \end{align}
\end{replemma}
\begin{proof}
In this proof we omit $\pi$ in the subscript of $\Delta_\pi(y)$ and in the superscript of $\mu^\pi_{y,y',a}$ for readability.

To calculate $G^\pi \circ f_\Delta^\pi(q, y)$,
we begin by applying \cref{lem:generator-ak}:
\begin{align}
    G^\pi \circ f_\Delta^\pi(q,y)
            &= \lambda (q - \Delta(y) + \frac{1}{2}) \label{eq:lam-term-main}\\
            &+\sum_{y', a} \mu_{y, y', a} \left(\frac{1}{2}\left((q-a)^+ - \Delta(y')\right)^2 - \frac{1}{2}\left(q - \Delta(y) \right)^2  \right). \label{eq:mu-term-main}
\end{align}
Recall that the unused service $u=\mathbbm{1}\{q =0 \land a=1\}$, so $(q-a)^+ = q-a+u$. We can decompose \eqref{eq:mu-term-main} into two terms, with and without $q$:
\begin{align}
    \eqref{eq:mu-term-main} &= 
    \label{eq:q-term-main}
    q \sum_{y', a} \mu_{y, y', a}  \left(-a + u - \Delta(y') + \Delta(y)\right) \\
    \nonumber
    &
    +   \sum_{y', a} \mu_{y, y', a} \left(\frac{1}{2} (-a + u - \Delta(y'))^2 - \frac{1}{2} \Delta(y)^2 \right).
\end{align}
The coefficient of $q$ in \eqref{eq:q-term-main} can be simplified considerably using \cref{lem:generator-delta}.
\begin{align}
\nonumber
    &\mspace{20mu} 
    \sum_{y', a} \mu_{y, y', a}  \left(-a + u - \Delta(y') + \Delta(y)\right)\\
\nonumber
    &= \sum_{y', a} \mu_{y, y', a} (-a) + \sum_{y', a} \mu_{y, y', a} u - \sum_{y', a} \mu_{y, y', a} (\Delta(y') - \Delta(y)) \\
\nonumber
    &= -\mu_{y,\cdot,1} - G^\ak \circ \Delta(y) + \sum_{y', a} \mu_{y, y', a} u \\
\nonumber
    &= -\mu_{y,\cdot,1} - (\comppi - \mu^\ak_{y,\cdot,1}) + \sum_{y', a} \mu_{y, y', a} u \\
\nonumber
    &= -\comppi + \sum_{y', a} \mu_{y, y', a} u .
\end{align}

Note that either $u=0$ or $q=0$,
because new jobs are only generated if the queue is empty.
As a result, $qu = 0$.
We can therefore further simplify the $q$-term in \eqref{eq:q-term-main}:
\begin{align}
    \label{eq:q-term-simple}
    q(\sum_{y', a} \mu_{y, y', a} u - \comppi)
    = - q \comppi
\end{align}
Substituting \eqref{eq:q-term-simple} into \eqref{eq:q-term-main}, \eqref{eq:q-term-main} into \eqref{eq:mu-term-main}, and performing some rearrangement, we find that 
\begin{align*}
    G^\pi \circ f^\pi_{\Delta}(q, y) &= (\lambda - \comppi)q  - \lambda \Delta(y) + \frac{1}{2}\lambda+ \sum_{y', a} \mu_{y, y', a} \left(\frac{1}{2} (-a + u - \Delta(y'))^2 - \frac{1}{2} \Delta(y)^2 \right).
    \qedhere
\end{align*}
\end{proof}

\begin{lemma}\label{lem:yd-distribution}
    In the MMSR-$\pi$ system, the departure average distribution $Y^\pi_d$ is given by
    \begin{equation}
        \frac{1}{\comppi} \Ep_{y\sim Y^\pi}[\mu^\pi_{y,y',1}] = \Prob(Y^\pi_d = y').
    \end{equation}
\end{lemma}

\begin{proof}

We will show that
\begin{align*}
    \Prob(Y^\pi_d = y') = \frac{1}{\comp_\pi} \sum_{y} \Prob(Y^\pi = y) \mu_{y, y', 1}.
\end{align*}

As an intermediate step, let $Y^\pi_{DTMC}$ be the transition-average steady state of the Markov chain $\pi$.
$\Prob(Y^\pi_{DTMC} = y)$ is the fraction of state-visits that are visits to $y$, in the embedded DTMC of $\pi$.

Let $\mu_{y,\cdot,\cdot}$ be the total transition rate out of state $y$:
\begin{align*}
    \mu_{y,\cdot,\cdot} = \sum_{y', a} \mu_{y,y',a}.
\end{align*}

Note that the CTMC that controls $Y^\pi$ and the DTMC that controls $Y^\pi_{DTMC}$ visit the same states in the same order, but that $Y^\pi$ stays in state $y$ for $Exp(\mu_{y,\cdot,\cdot})$ time for each visit.
As a result,
\begin{align*}
    \Prob(Y^\pi = y) = a^\pi \Prob(Y^\pi_{DTMC} = y) \frac{1}{\mu_{y,\cdot,\cdot}}.
\end{align*}
where $a^\pi$ is a normalization constant.
Specifically, $a^\pi$ is the long-term transition rate,
which can be calculated as the reciprocal of the average time per visit to a state: 
\begin{align*}
    a^\pi = \left( \sum_y \Prob(Y^\pi_{DTMC} = y) \frac{1}{\mu_{y,\cdot,\cdot}} \right)^{-1}.
\end{align*}

From $Y^\pi_{DTMC}$, we can calculate the fraction of transitions that move from a generic state $y$ to another generic state $y'$ via a completion.
Call this fraction $p_{y \to y',1}$:
\begin{align*}
    p_{y \to y',1} = \Prob(Y^\pi_{DTMC} = y) \frac{\mu_{y,y',1}}{\mu_{y,\cdot,\cdot}}.
\end{align*}
Summing over all initial states $y$, we can find the fraction of transitions that are completions which result in the state $y'$:
\begin{align*}
    p_{\cdot \to y',1} = \sum_{y} \Prob(Y^\pi_{DTMC} = y) \frac{\mu_{y,y',1}}{\mu_{y,\cdot,\cdot}}.
\end{align*}
Let $b^\pi$ be the overall fraction of transitions that are completions.
Conditioning on the transition into state $y'$ being a completion, we find that the probability that a generic completion results in state $y'$ is
\begin{align*}
    \Prob(Y^\pi_d = y') = \frac{p_{\cdot \to y',1}}{b^\pi}.
\end{align*}
Combining all of the above equations, we find that
\begin{align*}
    \Prob(Y^\pi_d = y')
    &= \frac{1}{b^\pi} \sum_{y} \Prob(Y^\pi_{DTMC} = y) \frac{\mu_{y,y',1}}{\mu_{y,\cdot,\cdot}} \\
    &= \frac{1}{b^\pi} \sum_{y} \frac{1}{a^\pi} \Prob(Y^\pi = y) \mu_{y,\cdot,\cdot} \frac{\mu_{y,y',1}}{\mu_{y,\cdot,\cdot}} \\
    &= \frac{1}{a^\pi b^\pi} \sum_{y} \Prob(Y^\pi = y) \mu_{y,y',1}.
\end{align*}

Recall that $a^\pi$ is the long-term transition rate, and that  $b^\pi$ is the fraction of transitions that are completions.
Thus, $a^\pi b^\pi$ is the long-term completion rate $X^\pi = \lambda^*_\pi$.
\end{proof}

\section{Lemmas about $G^{\msj}$}
\label{app:generator-msj-lemmas}
\begin{lemma}
    \label{lem:generator-msj}
    For any $f(q,y)$ which is a real-valued function of the state of the MSJ system,
    \begin{align}
        G^\msj \circ f(q,y) &=  \lambda \left(f(q+1, y) - f(q, y) \right) \indic_{\{y\in\ysak\}} \label{eq:Gmsj-f-term1}\\
        &+ \mathbbm{1}_{q=0, y \not\in \ysak}  \lambda \sum_{i \in S} p_i (f(0, y \cdot i)-f(0, y)) \label{eq:Gmsj-f-term2}\\
        &+ \indic_{q>0} \sum_{\substack{y'\in\ysak,\\ a\in\{0,1\}}} \mu^\ak_{y, y', a} \left(f((q-a)^+, y') - f(q, y)\right)
        \label{eq:Gmsj-f-term3} \\
        &+ \indic_{q=0} \sum_{\substack{y'\in\ysmsj,\\ a\in\{0,1\}}} \mu^\msj_{y, y', a, 0} \left(f((q-a)^+, y') - f(q, y)\right). \label{eq:Gmsj-f-term4}
    \end{align}
\end{lemma}
\begin{proof}
    Recall the definition of the generator 
    \[
        G^\msj \circ f(q, y) = \lim_{t \to 0} \frac{1}{t} \Ep[f(Q^\msj(t), Y^\msj(t)) - f(q, y) | Q^\msj(0) = q, Y^\msj(0)=y],
    \]
    which can be interpreted as the instantaneous rate of change of the function $f(Q^\msj(t), Y^\msj(t))$ when $(Q^\msj(t), Y^\msj(t))$ is initialized in $(q, y)$. Note that $(Q^\msj(t), Y^\msj(t))$ can change either due to an arrival event, or a transition event of the front state. An arrival event happens with rate $\lambda$, and its effect depends on whether $y\in \ysak$: if $y\in \ysak$, there are $k$ jobs in the front, so $Q^\msj(t)$ changes from $q$ to $q+1$, $Y^\msj(t)$ remains unchanged; if $y\notin \ysak$, there are strictly fewer than $k$ jobs in the front, so $Q^\msj(t)$ remains zero after the arrival, and $Y^\msj(t)$ changes from $y$ to $y \cdot i$ with probability $p_i$ (append a fresh job in state $i$ to the front state with probability $p_i$). Therefore, arrival events contribute
    \begin{align*}
         &\lambda \left(f(q+1, y) - f(q, y) \right) \indic_{\{y\in\ysak\}} \\
        &+ \mathbbm{1}_{q=0, y \not\in \ysak}  \lambda \sum_{i \in S} p_i (f(0, y \cdot i)-f(0, y))
    \end{align*}
    to $G^\msj \circ f(q, y)$, which are the terms in \eqref{eq:Gmsj-f-term1} and \eqref{eq:Gmsj-f-term2} in the lemma statement. As for the transition events of the front, a transition from state $y$ to state $y'$ accompanied by $a$ completions causes $(Q^\msj(t), Y^\msj(t))$ to change from $(q, y)$ to $((q-a)^+, y')$. Such a transition happens with the rate $\mu^\msj_{y, y', a, 1} = \mu^\ak_{y,y',a}$ if $q > 0$ and $y' \in \ysak$, and happens with rate $\mu^\msj_{y, y', a, 0}$ if $q = 0$. Therefore, the transition events of the front contribute 
    \begin{align*}
        &\indic_{q>0} \sum_{\substack{y'\in\ysak,\\ a\in\{0,1\}}} \mu^\ak_{y, y', a} \left(f((q-a)^+, y') - f(q, y)\right)
        \\
        &+ \indic_{q=0} \sum_{\substack{y'\in\ysmsj,\\ a\in\{0,1\}}} \mu^\msj_{y, y', a, 0} \left(f((q-a)^+, y') - f(q, y)\right)
    \end{align*}
    to $G^\msj \circ f(q, y)$, which are the terms in \eqref{eq:Gmsj-f-term3} and \eqref{eq:Gmsj-f-term4} in the lemma statement. 
\end{proof}

\begin{replemma}{lem:generator-f-msj}
\begin{align}
    G^\msj \circ f^\msj_\Delta(q, y) = \mathbbm{1}_{q > 0} G^\ak \circ f_\Delta^\ak(q, y) + \mathbbm{1}_{q=0} O_\lambda(1)
\end{align}
\end{replemma}
\begin{proof}
Let us begin by using \cref{lem:generator-ak,lem:generator-msj} to give expressions for $G^\msj \circ f^\msj_\Delta(q, y)$ and $G^\ak \circ f_\Delta^\ak(q, y)$.

Note that whenever $q>0$,
$G^\msj \circ f^\msj_\Delta(q, y)$ is identical to $G^\ak \circ f_\Delta^\ak(q, y)$,
because the two systems have the same transitions
and because $f^\msj_\Delta(q, y)$ and $f_\Delta^\ak(q, y)$ are identical.

Note also that whenever $q=0$,
both $G^\msj f^\msj_\Delta(q, y)$ and $G^\ak \circ f_\Delta^\ak(q, y)$
are $O_\lambda(1)$, because $\Delta(y)$ is bounded by a constant for all $y$,
because $\ysmsj$ is finite.

As a result,
\begin{align*}
    G^\msj \circ f^\msj_\Delta(q, y) = \mathbbm{1}_{q > 0} G^\ak \circ f_\Delta^\ak(q, y)  + \mathbbm{1}_{q=0} O_\lambda(1).
    \qquad
 \qedhere
 \end{align*}
\end{proof}

\section{At-least-$k$ busy period}
We also prove a lemma about busy periods in the At-least-$k$ system.
Define a busy period to begin when the back length $q^\ak$ in the At-least-$k$ system
transitions from 0 to 1,
and to end when the back length next returns to 0.
Let $B^\ak$ be a random variable representing the length of a busy period in the At-least-$k$ system
in stationarity. 
\begin{lemma}\label{lem:busy-period-ak}
    In the At-least-$k$ system, for all $\lambda < \comp$
    \begin{align}
        \Ep[B^\ak] = \Omega_\lambda \left(\frac{1}{1-\lambda/\lambda^*} \right).
    \end{align}
\end{lemma}
\begin{proof}
    In this proof we omit $\ak$ in the subscript of $\mu^\ak_{y,y',a}$ for readability. 
    
    To prove \cref{lem:busy-period-ak}, it suffices to show that
    $\Prob(Q^{\ak}=0) = O_{\lambda}(1-\frac{\lambda}{\comp})$, and that the non-busy periods
    (periods when $Q^\ak=0$)
    have expected duration $\Omega_\lambda(1)$.
    The latter follows from the fact that all transitions have expected duration $\Omega_\lambda(1)$. 
    
    To prove the former, let $u(q^\ak, y^\ak)$ be the rate at which new jobs are generated
    due to completions in a particular state $(q^\ak, y^\ak)$ of the At-least-$k$ system.
    Note that $u(q^\ak, y^\ak)$ is positive only if $q^\ak=0$.
    The time-average value of $u(q^\ak, y^\ak)$
    is the difference between the completion rate of the system and the Poisson arrival rate,
    because in steady state the total completion rate and total arrival rate must match.
    Thus,
    \begin{align}
        \Ep[u(Q^{\ak}, Y^{\ak})] = \comp - \lambda. 
    \end{align}
    Note that $u(q, y) = \mu_{y,\cdot,1}\indic_{\{q=0\}}$, so 
    \begin{align*}
        \Ep[\mu_{Y^{\ak},\cdot,1}\indic_{\{Q^{\ak}=0\}}] = \comp - \lambda.
    \end{align*}
    Note that
    \begin{align}
        \Prob(Q^{\ak} = 0) = \frac{\Ep[\mu_{Y^{\ak},\cdot,1}\indic_{\{Q^{\ak}=0\}}]}{\Ep[\mu_{Y^{\ak},\cdot,1} | Q^{\ak}=0]} = \frac{\comp-\lambda}{\Ep[\mu_{Y^{\ak},\cdot,1} | Q^{\ak}=0]}.
    \end{align}
    It therefore suffices to show that there exists a constant $c > 0$ not dependent on $\lambda$ such that $\Ep[\mu_{Y^{\ak},\cdot,1} | Q^{\ak}=0]\ge c$.

    From an arbitrary state $y^\ak$ with $q^\ak=0$,
    the distribution of time until a completion next occurs does not depend on $\lambda$.
    Consider the probability of a completion happening in the next second,
    with no arrivals happening before that completion.
    This probability is nonzero, and only dependent only $\lambda$ via the arrival process.
    The probability can be lower bounded away from zero by substituting a Poisson$(\comp)$ process instead.
    We can thus lower bound the completion rate over the next second with an empty back away from 0.
    This therefore provides a lower bound on the completion rate conditional on the back being empty,
    $\Ep[\mu_{Y^{\ak},\cdot,1} | Q^{\ak}=0]$, as desired.
    Calling that lower bound $c$, we have:
    \begin{align}
        \Prob(Q^{\ak} = 0) = \frac{\Ep[\mu_{Y^{\ak},\cdot,1}\indic_{\{Q^{\ak}=0\}}]}{\Ep[\mu_{Y^{\ak},\cdot,1} | Q^{\ak}=0]} \le \frac{\comp - \lambda}{c} = O_{\lambda}\left(1-\frac{\lambda}{\comp} \right).
    \end{align}

    This completes the proof.
\end{proof}

\section{Coupling Lemmas}
\label{app:coupling-lemmas}

Let us restate the coupling between the At-least-$k$ and MSJ systems. We let the arrivals of the two systems happen at the same time. We couple the transitions of their front states based on their joint state $(q^\msj, y^\msj, q^\ak, y^\ak)$. If $y^\msj=y^\ak$, $q^\msj>0$, and $q^\ak>0$, the completions happen at the same time in both systems, the same jobs complete, the same job phase transitions occur, and the jobs entering the front are the same.
We call the two systems ``merged'' during such a time period. Note that under this coupling, if the two systems become merged,
they will stay merged until $q^\msj=0$ or $q^\ak=0$.
If the systems are not merged, the two systems have independent completions and phase transitions, and independently sampled jobs.

The two systems transition according to synchronized Poisson timers whenever
they are merged,
and independent Poisson timers otherwise.
Because all transitions are exponentially distributed,
this poses no obstacle to the coupling.

\begin{replemma}{lem:quick-merge}[Quick merge]
    From any joint MSJ, At-least-$k$ state, for any $\epsilon > 0$, under the coupling above,
    the expected time until $y^{\msj}=y^{\ak}$, $q^{\msj} \ge k+1$, and $q^{\ak} \ge k+1$
    is at most $m_1(\epsilon)$ for some $m_1(\epsilon)$
    independent of the arrival rate $\lambda$ and initial joint states,
    given that $\lambda \in [\epsilon, \lambda^*)$.
\end{replemma}

\begin{proof}

    We call the period of time until $y^{\msj}=y^{\ak}$,
    $q^{\msj} \ge k+1$, and $q^{\ak} \ge k+1$
    the ``bad period.'' 
    We wish to show that the expected length of the bad period is upper bounded by some constant $m_1(\epsilon)$ for all $\lambda$ such that $\lambda \in [\epsilon, \lambda^*)$.
    
    Consider the possibility that the following sequence of events occurs:
    over a period of 1/2 second, at least $2k+1$ jobs arrive.
    Then, over another 1/2 second, $k$ completions occur in each of the MSJ and At-least-$k$ systems, which is sufficient to clear out every job initially present in the fronts and replace them with freshly sampled jobs.
    Finally, the sampled jobs in the fronts of the two systems are the same, in the same order.
    After this sequence of events, $y^\msj = y^\ak$, $q^{\msj} \ge k+1$, and $q^{\ak} \ge k+1$, which ends the bad period. 
    
    Recall that as long as the front states of the two systems are distinct,
    their completions are independent.
    As a result, the probability of this sequence of events is positive,
    for any $\lambda > 0$ and for any initial states $y^{\msj},y^{\ak}$.
    We call the probability of this sequence of events $\pgood(\lambda, y^{\msj},y^{\ak})$. 
    
    Moreover, $\pgood(\lambda, y^{\msj},y^{\ak})$ is monotonically increasing in $\lambda$, 
    as $\lambda$ only affects the probability that at least $2k+1$ jobs arrive
    in the first half second.

    Therefore, the least value of $\pgood(\lambda, y^{\msj},y^{\ak})$ is achieved when $\lambda = \epsilon$. 
    Because there are only finitely many possible front states $y^{\msj} \in \ysmsj,y^{\ak}\in \ysak$, 
    there must be some lowest value of $\pgood(\epsilon, y^{\msj},y^{\ak})$. 
    We call this value $\pgood^*(\epsilon)$.
    Note that for all $\lambda \ge \epsilon$ and for all $y^{\msj} \in \ysmsj,y^{\ak} \in \ysak,$
    \begin{align}
        \pgood(\lambda, y^{\msj},y^{\ak}) \ge \pgood^*(\epsilon) > 0.
    \end{align}
    In the first second, there is at least a $\pgood^*(\epsilon)$
    chance of the desired sequence of events happening and the bad period completing.
    In the next second, the same is true.
    In general, the time until the bad period completes is upper bounded by
    a geometric distribution with completion probability $\pgood^*(\epsilon)$.
    Taking $m_1(\epsilon) = 1/\pgood^*(\epsilon)$, the mean time until the bad period completes
    is upper bounded by $m_1(\epsilon)$, which is independent of the arrival rate $\lambda$ and initial joint states, as desired.     
\end{proof}

\begin{replemma}{lem:long-busy}[Long merged period]
    From any joint MSJ, At-least-$k$ state such that
    $y^{\msj}=y^{\ak}$, $q^{\msj} \ge k+1$, and $q^{\ak} \ge k+1$, the expected time until $q^{\msj}=0$,
    $q^{\ak}=0$, or $y^{\msj} \neq y^{\ak}$,
    is at least $\frac{m_2}{1-\lambda/\lambda^*}$ for some $m_2$ independent of the arrival rate $\lambda$ and initial joint states, given that $\lambda < \lambda^*$. 
\end{replemma}
Note that the time until $q^{\msj}=0$ or $q^{\ak}=0$ is a lower bound on the time until $y^{\msj} \neq y^{\ak}$.

\begin{proof}
    In this proof we omit $\ak$ in the subscript of $\mu^\ak_{y,y',a}$ for readability.
    
    Let's call the period of interest the ``good period.''
    Note that throughout the good period, $y^{\msj}=y^{\ak}$.
    Let us introduce a new lower-bounding MSJ system, $M'$, beginning in a general state
    $y^{M'}=y^\msj=y^\ak$ and beginning with $q^{M'} = k+1$.
    Let us define a coupling between $M'$ and the original MSJ and At-least-$k$ systems in the same synchronized/independent fashion defined at the start of \cref{sec:coupling}.
    As a result,
    for all time until $q^{M'} = 0$,
    $y^{M'}=y^\msj=y^\ak$, and
    $q^{M'} \le q^\msj$, and $q^{M'} \le q^\ak$.
    In particular, the duration until $q^{M'} = 0$
    is a lower bound on the length of the good period.

    Let us set up a new coupled system, $M''$.
    The $M''$ system is an At-least-$k$ system initialized in a specific front state distribution to be specified later
    and with $q^{M''}=1$.
    Let us define a coupling between the $M'$ and $M''$ systems in the same synchronized/independent fashion defined at the start of \cref{sec:coupling}.
    Note however that $M''$ is a new system, distinct from all of the previous systems. 

    Let $B^{M'}$ be the length of the first busy period of the $M'$ system, which is the time in $M'$ until $q^{M'} = 0$; similarly, let $B^{M''}$ be the length of the first busy period of the $M''$ system. We want to show that 
    \begin{align}
        \Ep[B^{M'}] &\geq m_3 \Ep[B^{M''}], \label{eq:long-bp-proof:msj-longer-bp-than-ak}\\
        \Ep[B^{M''}] &\geq \frac{m_4}{1-\lambda/\lambda^*}. \label{eq:long-bp-proof:ak-long-bp}
    \end{align}
    for some positive numbers $m_3$ and $m_4$ independent of the arrival rate $\lambda$ and the initial front state of the $M'$ system $y^{M'}$. 

    We will choose the front state distribution of the $M''$
    system in order to guarantee that \eqref{eq:long-bp-proof:ak-long-bp} holds. 
    To do so, we will make use of \cref{lem:busy-period-ak}, which states that the At-least-$k$ system has long busy periods:
    \begin{align}
        \Ep[B^\ak] = \Omega_\lambda \left(\frac{1}{1-\lambda/\comp} \right).
    \end{align}
    
    Let $Y^{\ak-BP}$ denote the long-term-average 
    distribution of the front state in the At-least-$k$ system
    at the start of a busy period. 
    We let the initial state distribution of the $M''$ system be $y^{M''} \sim Y^{\ak-BP}$ and $q^{M''}=1$. As a result, $\Ep[B^{M''}] = \Ep[B^\ak]$, the expected busy period length of the At-least-$k$ system. By \cref{lem:busy-period-ak}, we have $\Ep[B^{M''}] \geq \frac{m_4}{1-\lambda/\lambda^*}$ for some positive number $m_4$ independent of $\lambda$ and $y^{M'}$. 

    Now, we wish to show \eqref{eq:long-bp-proof:msj-longer-bp-than-ak}: that the length of the first busy period in $M'$,
    initialized in an arbitrary initial front state $y^{M'}$ and $q^{M'}=k+1$,
    is also long in expectation.
    
    To prove this, let us introduce a very fast Poisson process with a rate $\mu^*$ given by 
    \[
        \mu^* = \comp + \max_{y \in \ysak} \sum_{y', a} \mu_{y, y', a}. 
    \]
    Note that $\mu^*$ is at least as fast as the transition rate of $M''$ in any state, and $\mu^*$ is independent of $\lambda$.
    Let us define a coupling between the Poisson($\mu^*$) process and the $M''$ system.
    Transitions in the $M''$ system only occur when the Poisson($\mu^*$) increment occurs,
    where with some probability sampled on each Poisson increment
    a transition happens, and otherwise no transition occurs. In state $y$, a transition happens with probability
    \begin{align*}
        \frac{\lambda + \sum_{y',a} \mu_{y,y',a}}{\mu^*}.
    \end{align*}
    Note that this probability is always less than 1, by the definition of $\mu^*$.
    
    To lower bound $\Ep[B^{M'}]$, the expected busy period length in the $M'$ system,
    let us consider $\Ep[B^{M'}\indic_{\{A_1 \wedge A_2\}}]$,
    where $A_1$ and $A_2$ are the following two events:
    \begin{enumerate}
        \item Event $A_1$: the first increment of the Poisson($\mu^*$) process takes at least 1 second.
        \item Event $A_2$: during the first second $M'$ has exactly $k$ completions,
        after each of which the job entering the front of the $M'$ system is sampled to have the same server need as the corresponding job of the $M''$ system,
        and then all the jobs transition to the same phase as in the $M''$ system.
        At the end of the first second,
        $M'$ and $M''$ have identical front states $y$ and back lengths $q=1$ after exactly $k$ completions in the $M'$ system.
    \end{enumerate}
    First, note that 
    \begin{align}
        \Ep[B^{M'}] \ge \Ep[B^{M'}\indic_{\{A_1 \wedge A_2\}}] = \Ep[B^{M'} |A_1 \wedge A_2 ] \Prob(A_1 \wedge A_2).
    \end{align}
    Note that $\Prob(A_1 \wedge A_2)$ is lower bounded by a positive constant
    for every $\lambda$ such that $\epsilon \leq \lambda \leq \comp$, so we can focus on $\Ep[B^{M'} |A_1 \wedge A_2 ]$.
    
    Note that if events $A_1$ and $A_2$ occur,
    the $M'$ and $M''$ systems have the same busy period length,
    because after 1 second, the two systems have identical states. Specifically, both systems become empty at the same time, which is the first time after each is initialized when each becomes empty.
    
    As a result,
    \begin{align}
        \Ep[B^{M'} |A_1 \wedge A_2 ] = \Ep[B^{M''} |A_1 \wedge A_2 ].
    \end{align}
    
    Note that Event $A_2$ is conditionally independent of the behavior of the $M''$ system,
    given that Event $A_1$ occurs.
    As a result, 
    \begin{align}
        \Ep[B^{M''} |A_1 \wedge A_2] = \Ep[B^{M''} |A_1].
    \end{align}
    
    Notice that event $A_1$ is independent of the state of $M''$.
    Conditioning on the event $A_1$ merely increases the time of the first transition in $M''$,
    without altering the future updates of $M''$ at all.
    As a result,
    \begin{align}
        \Ep[B^{M''} |A_1] \ge \Ep[B^{M''}].
    \end{align}
    
    Thus, $\Ep[B^{M'}]$ is lower bounded by a constant multiple of $\Ep[B^{M''}]$.
    Recall that by construction, $\Ep[B^{M''}] =\Omega_\lambda(\frac{1}{1-\lambda/\comp})$. 
    Combining \eqref{eq:long-bp-proof:msj-longer-bp-than-ak} and \eqref{eq:long-bp-proof:ak-long-bp} and letting $m_2 = m_3 m_4$, we get the desired lower bound on the expected length of the good period. 
\end{proof}

\section{Extensions of the Multiserver-job model}
\label{app:additional-msj}
\subsection{Nontrival scheduling policies: Backfilling}
A common family of MSJ scheduling policies in practice are \emph{backfilling} policies \cite{carastan_one_2019,wang_application_2009,srinivasan_characterization_2002}.
Under a backfilling policy, the scheduler begins by placing jobs into service in arrival order, as in the FCFS policy.
However, once a job is encountered which does not fit in the available servers, additional jobs are considered for service.
By doing so, the stability region and mean response time are improved relative to FCFS, though it is unclear whether the full stability region can be achieved \cite{grosof_serverfilling_2023}.
Some backfilling policies give rise to finite skip models,
and can thus be handled by the RESET technique.

As an example, consider the ``First Fit-$k$'' policy:
The scheduler iterates through the first $k$ jobs in arrival order, checking for each job whether it can be served in the available servers. Each job that fits is served.
This policy only serves jobs among the $k$ oldest in arrival order,
so it can be handled by the RESET technique.

Beyond backfilling policies, more advanced packing policies can also be considered.
For instance, for small $k$ the scheduler could simply search over all subsets of the $k$ oldest jobs and serve the subset with maximal total server need $\le k$.
This policy is also finite skip, and the RESET technique also applies.

\subsection{Changing server need during service}

The standard MSJ model assumes that jobs require a fixed service need throughout their time in service. However, in some settings, jobs may require  a varying number of servers.
For example, consider a fork-join model with simultaneous start.
Suppose that each job is made of some number of tasks,
each with independent duration, and each requiring 1 server.
As the tasks complete, the server need of the job as a whole diminishes,
freeing up space for other jobs to run.
This setting still gives rise to a finite-skip model,
and poses no difficulty to our RESET technique.

Another natural setting in which server needs change over time is the directed acyclic graph (DAG) setting, in which jobs are broken up into small segments of work, with potentially complex dependencies between segments.
The DAG scheduling literature often focuses on scheduling the segments of an individual DAG job.
It is natural to consider a scheduling setting where many DAG jobs arrive over time.
Holding the DAG scheduling policy constant, this model effectively gives rise to a MSJ model where server needs can vary over time, and potentially vary dynamically in response to the service conditions.
As long as the high-level scheduling policy deciding which DAG jobs to run is finite-skip,
the model as a whole is finite-skip, and our RESET technique can characterize its asymptotic mean response time.

\subsection{Multidimensional resource constraints}
The standard MSJ model considers a single constrained resource.
However, computing jobs are often constrained by a variety of resources,
such as CPU, GPU, other accelerators, memory bandwidth, cache capacity, network bandwidth, etc.
Such multidimensional resource constraints are often considered in the VM scheduling literature.
In that literature, only stability results are known. Our RESET technique thus gives the first characterization of asymptotic mean response time in that setting.

\subsection{Heterogeneous servers}
In the standard MSJ model, all servers are identical. However, it is also important to consider settings where different kinds of servers are available, which can provide different amounts of resources.
One can also consider jobs that need to be served at a particular server or set of servers,
such as a job that processes data stored at that server.
In a multidimensional resource setting, some servers may also provide different resources, such as a GPU-heavy or CPU-heavy server.
All of these extensions are compatible with the RESET technique.

\subsection{Turning off idle servers}
To improve energy efficiency, it may be preferable to turn off idle servers.
Idle servers consume nearly as much energy as active servers.
However, turned-off servers take some time to restart.
It is important to characterize the impact of this start-up delay on mean response time
to understand the tradeoff inherent in turning off idle servers.
The process of turning off and on servers can be incorporated into a finite-skip model,
because there are a finite number of possible states that the servers can be in.
As a result, our RESET technique can provide a characterization of mean response time.

\subsection{Preemption overheads}
The FCFS policy never preempts any jobs. Prior work has studied settings with unlimited preemption. However, practical settings often allow only a limited subset of jobs to be preempted, and jobs may incur an overhead when preemption occurs. This overhead corresponds to the time necessary to snapshot a job in service, and for the new job to be transferred onto the freed servers.
Models with preemption overheads have only recently begun to be analyzed in the M/G/1 setting \cite{peng_exact_2022},
with no mean response time analysis known in the one-server-per-job multiserver model,
much less the multiserver-job model.
Preemption overheads can be modeled with a finite-number of additional states, marking the corresponding servers as undergoing preemption.
As a result, our RESET technique can provide a characterization of mean response time.



\section{Empirical correlation between $\Delta(Y_d^\sat)$ and convergence rate}
\label{app:delta-convergence}

\begin{figure}
    \centering
    \includegraphics{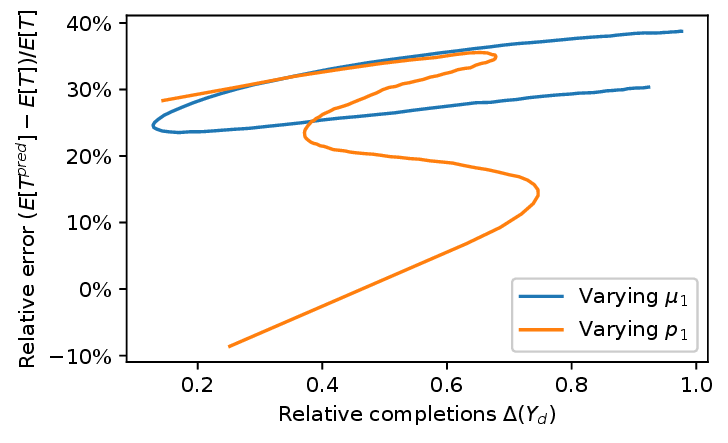}
    \caption{Empirical relationship between $\Delta(Y_d^\sat)$ and the relative error between our prediction of mean response time and the true value,
    in two parameterized workload settings, both with $k=5$ servers and server needs either 1 or 5.
    We alternately parameterize $\mu_1$, the completion rate of 1-server jobs, and $p_1$, the fraction of 1-server jobs. Load $\lambda/\comp = 0.8$.
    Simulated $10^8$ arrivals.}
    \label{fig:delta-convergence}
\end{figure}
As discussed in \cref{sec:empirical},
we have empirically noticed a correlation between large $\Delta(Y_d^\sat)$ values
and slower convergence rates of our predicted value of $E[T]$ to the exact value,
as $\lambda \to \comp$.
Our predicted value of mean response time is:
\begin{align*}
    E[T^{pred}] = \frac{1}{\comp} \frac{\Delta(Y_d^\sat) + 1}{1-\lambda/\comp}.
\end{align*}
We prove in \cref{thm:msj-response-time} that $E[T^{pred}] - E[T] = O_\lambda(1)$,
ensuring that the two values reliably converge in the $\lambda \to \comp$ limit
in all settings, as illustrated in \cref{sec:empirical}.

In this section, we further investigate this correlation by comparing $\Delta(Y_d^\sat)$ with the relative error $\frac{E[T^{pred}]-E[T]}{E[T]}$
for a pair of parameterized sequences of workload settings.
The setting we investigate has $k=5$ servers, with jobs having server need either 1 or 5.
We set the arrival rate $\lambda = 0.8\comp$, using $80\%$ of the stability region.

We separately parameterize the fraction of 1-server jobs present, as well as the duration of 1-server jobs.
First, we vary $p_1$, the fraction of 1-server jobs from $1\%$ to $99\%$, in $1\%$ increments,
while setting 1-server jobs to have duration $Exp(1/5)$.
Second, we set the duration of 1-server jobs to be $Exp(\mu_1)$, with $\mu_1$ ranging from $0.01$ to $100$ in 100 evenly multiplicatively-spaced increments,
while setting $50\%$ of jobs to have each server need.
In both cases, we set the 5-server jobs to have duration $Exp(1)$.

We plot the behavior of these two settings in \cref{fig:delta-convergence},
comparing the $\Delta(Y_d^\sat)$, the relative completion in the departure-average state of the saturated system,
against the relative error $\frac{E[T^{pred}] - E[T]}{E[T]}$.
The empirical results show a significant correlation between $\Delta(Y_d^\sat)$ and the relative error $\frac{E[T^{pred}] - E[T]}{E[T]}$ in the case of parameterized $\mu_1$ ($R^2 = 0.526$), but no significant correlation in the case of parameterized $p_1$ ($R^2=0.005$).
This indicates that the correlation observed in \cref{sec:empirical} may exist in some settings, but is not robust or reliable.
Further investigation will be needed to better understand this correlation.

\end{document}

%% file: coupling-lemmas-proofs.tex
\subsection{Coupling between At-least-$k$ and MSJ}
\label{sec:coupling}

To show that the Ak system and the MSJ system have identical asymptotic mean response time, we define the following coupling of the two systems. We let the arrivals of the two systems happen at the same time. We couple the transitions of their front states based on their joint state $(q^\msj, y^\msj, q^\ak, y^\ak)$. If $y^\msj=y^\ak$, $q^\msj>0$, and $q^\ak>0$, the completions happen at the same time in both systems, the same jobs complete, the same job phase transitions occur, and the jobs entering the front are the same.
We call the two systems ``merged'' during such a time period. Note that under this coupling, if the two systems become merged,
they will stay merged until $q^\msj=0$ or $q^\ak=0$.
If the systems are not merged, the two systems have independent completions and phase transitions, and independently sampled jobs.

The two systems transition according to synchronized Poisson timers whenever
they are merged,
and independent Poisson timers otherwise.
Because all transitions are exponentially distributed,
this poses no obstacle to the coupling.

We want to show that under this coupling,
the two systems spend almost all of their time merged,
in the limit as $\lambda \to \lambda^*$.
Specifically, we will show that the fraction of time in which the two systems are \emph{unmerged} is $O_\lambda(1-\frac{\lambda}{\lambda^*})$.
This implies \cref{lem:tight-coupling},
which is the key lemma we need for our main RESET result, \cref{thm:msj-response-time}.

\begin{lemma}[Tight coupling]
    \label{lem:tight-coupling}
    In the MSJ system, for any $\lambda < \lambda^*$, we have the following two properties:
    \begin{enumerate}
    \item Property 1: $P(Q^{\msj}=0) = O_\lambda(1-\frac{\lambda}{\lambda^*})$.
    \item Property 2: $P(Y^{\msj} \neq Y^{\ak}) = O_\lambda(1-\frac{\lambda}{\lambda^*})$.
    \end{enumerate}
    where property 2 holds under the coupling in \cref{sec:coupling}.
\end{lemma}

To prove \cref{lem:tight-coupling},
we prove two key lemmas:
\begin{itemize}
    \item \cref{lem:quick-merge}: Whenever the two systems are unmerged, the expected time until the systems become merged is $O_\lambda(1)$.
    \item \cref{lem:long-busy}: Whenever the two systems are merged, the expected time for which they stay merged is $\Omega_\lambda(\frac{1}{1-\lambda/\lambda^*})$.
\end{itemize}

We then use a renewal-reward approach to prove \cref{lem:tight-coupling}.

\begin{lemma}[Quick merge]
    \label{lem:quick-merge}
    From any joint MSJ, Ak state, for any $\epsilon > 0$, under the coupling above,
    the expected time until $y^{\msj}=y^{\ak}$, $q^{\msj} \ge k+1$, and $q^{\ak} \ge k+1$
    is at most $m_1(\epsilon)$ for some $m_1(\epsilon)$
    independent of the arrival rate $\lambda$ and initial joint states,
    given that $\lambda \in [\epsilon, \lambda^*)$.
\end{lemma}

\begin{lemma}[Long merged period]
    \label{lem:long-busy}
    From any joint MSJ, Ak state such that
    $y^{\msj}=y^{\ak}$, $q^{\msj} \ge k+1$, and $q^{\ak} \ge k+1$, the expected time until $q^{\msj}=0$,
    $q^{\ak}=0$, or $y^{\msj} \neq y^{\ak}$,
    is at least $\frac{m_2}{1-\lambda/\lambda^*}$ for some $m_2$ independent of the arrival rate $\lambda$ and initial joint states, given that $\lambda < \lambda^*$. 
\end{lemma}
\proof[Proofs deferred to \ref{app:coupling-lemmas}]{~}

Using \cref{lem:quick-merge,lem:long-busy}, we can prove \cref{lem:tight-coupling}: 
\begin{proof}
    Let $\epsilon = \frac{\comp}{2}$.
    Note that if $\lambda < \epsilon$,
    the properties are trivial:
    $O_\lambda(1-\frac{\lambda}{\comp})\equiv O_\lambda(1)$, and probabilities are bounded.
    Therefore, we will focus on the case where $\lambda \ge \epsilon$,
    where we can apply \cref{lem:quick-merge,lem:long-busy}.
    
    Let us define a \textit{good period} to begin when $Y^{\msj}(t) = Y^{\ak}(t)$, $Q^{\msj}(t) \geq k+1$ and $Q^{\ak}(t) \geq k+1$, and end when $Q^{\msj}(t) = 0$ or $Q^{\ak}(t) = 0$. Let a \textit{bad period} be the time between two good periods.
    Note that throughout a good period, the front states are merged ($Y^\msj(t)=Y^\ak(t)$)
    and both queues are nonempty.

    To bound the fraction of time that the joint system is in a good period,
    we introduce the concept of a ``$y^*$-cycle."
    Let $y^*$ be an arbitrary state in $\ysak$.
    Let a $y^*$-cycle be a renewal cycle whose renewal points are moments when a bad period begins, and $Y^\msj(t)=Y^\ak(t)=y^*$, and $Q^\msj(t)=Q^\ak(t)=0$, for some designated state $y^*$.
    We will show that a $y^*$-cycle has finite mean time.
    Given that fact, we can apply renewal reward to derive the equations below:
    \begin{align}
        P(Q^{\msj} = 0) &= \frac{\Ep[\text{$Q^{\msj}(t)=0$ time per $y^*$-cycle}]}{\Ep[\text{total time per $y^*$-cycle}]}, \label{eq:property1-renewal-reward} \\ 
        P(Y^{\msj} \neq Y^{\ak}) &= \frac{\Ep[\text{$Y^{\msj}(t)\neq Y^{\ak}(t)$ time per $y^*$-cycle}]}{\Ep[\text{total time per $y^*$-cycle}]}.\label{eq:property2-renewal-reward}
    \end{align}
    Note that $Q^{\msj}(t)=0$ or $Y^{\msj}(t)\neq Y^{\ak}(t)$ only during a bad period, so the two probabilities in \eqref{eq:property1-renewal-reward} and \eqref{eq:property2-renewal-reward} are both bounded by the fraction of time spent in bad periods. 
    By Lemma~\ref{lem:quick-merge} and Lemma~\ref{lem:long-busy}, the expected length of a bad period is at most $m_1$ and the expected length of a good period is at least $\frac{m_2}{1-\lambda/\comp}$, conditioned on any initial joint state. 
    Let $Z$ be a random variable denoting the number of good periods in
    a $y^*$ cycle. Note that good and bad periods alternate.
    \begin{align*}
        \Ep[\text{total time per $y^*$-cycle}] &\geq \frac{m_2}{1-\lambda/\comp} \Ep[Z],\\
        \Ep[\text{bad period time per $y^*$-cycle}] &\leq m_1 \Ep[Z].
    \end{align*}
    If a $y^*$-cycle has finite mean time, then we also have $\Ep[Z] < \infty$ because each good period and bad period take a positive time. Plugging the above inequalities into \eqref{eq:property1-renewal-reward} and \eqref{eq:property2-renewal-reward},
    we derive Properties 1 and 2:
    \begin{align*}
        P(Q^{\msj} = 0) &\leq \frac{m_1}{m_2} \left(1-\frac{\lambda}{\comp} \right), \qquad
        P(Y^{\msj} \neq Y^{\ak}) \leq \frac{m_1}{m_2} \left(1-\frac{\lambda}{\comp} \right).
    \end{align*}

    It remains to show that a $y^*$-cycle has finite mean time. 
    We first use a Lyapunov argument to show that the joint states of the two systems return to a bounded set in a finite mean time. 
    Consider the Lyapunov function $f_\Delta^\msj(q^\msj, y^\msj) + f_\Delta^\ak(q^\ak, y^\ak)$. Its drift is:
    \begin{align*}
        G^{\msj,\ak} \circ \left(f_\Delta^\msj(q^\msj, y^\msj) + f_\Delta^\ak(q^\ak, y^\ak) \right) = G^\msj \circ f^\msj_\Delta(q^\msj, y^\msj) + G^\ak \circ f^\ak_\Delta(q^\ak, y^\ak).
    \end{align*}
    Applying \cref{lem:generator-f} to the Ak system,
    \[
       G^\ak \circ f^\ak_\Delta(q^\ak, y^\ak) = (\lambda - \comp) q^\ak + c_0(y^\ak,q^\ak),
    \]
    where $c_0(y, q)$ is defined in \cref{def:cs}.
    Note that $c_0(y, q)$ is a bounded function because $\Delta(y)$ is bounded, by \cref{lem:delta-exists}.
    Let $c_{\max}^\ak$ be the maximum of $c_0(y, q)$.
    For all $y^\ak, q^\ak$, 
    \[
        G^\ak \circ f^\ak_\Delta(q^\ak, y^\ak) \le (\lambda - \comp) q^\ak + c_{\max}^\ak.
    \]
    By similar reasoning, applying \cref{lem:generator-f-msj},
    there exists a $c_{\max}^\msj$ such that
    \begin{align*}
        G^\msj \circ f^\msj_\Delta(q^\msj, y^\msj) \le (\lambda - \comp) q^\ak + c_{\max}^\msj
    \end{align*}
    Let $c_{\max} = \max(c_{\max}^\ak,c_{\max}^\msj)$.
    Consider any $q^\ak \geq \frac{2 c_{\max}+1}{\comp - \lambda}$.
    Then for any $y^\ak$,
    \begin{align*}
        G^\ak \circ f^\ak_\Delta(q^\ak, y^\ak) \leq -c_{\max} -1.
    \end{align*}
    Similarly, for any $q^\msj \geq \frac{2 c_{\max}+1}{\comp - \lambda}$ and any $y^\msj$,
    \[
        G^\msj \circ f^\msj_\Delta(q^\msj, y^\msj) \leq -c_{\max}-1.
    \]
    
    Let $c_{\text{cap}}=\max\{\frac{2 c_{\max}+1}{\comp - \lambda}, k+1\}$.
    We define the bounded set $\s$ as
    \[
        \s = \left\{(q^\msj, q^\ak, y^\msj, y^\ak) \colon q^\msj \leq c_{\text{cap}}, q^\ak \leq c_{\text{cap}}\right\}.
    \]
    By the calculation above, outside $\s$, 
    \[
        G^\msj f^\msj_\Delta(q^\msj, y^\msj) + G^\ak f^\ak_\Delta(q^\ak, y^\ak) \leq -1.
    \]
    In particular, outside of $\s$, either $q^\msj > c_{\max}$ or $q^\ak > c_{\max}$,
    yielding a drift term $\le -c_{\max} -1$,
    outweighing the term where $q$ is small.
    Thus, by the Foster-Lyapunov theorem
    \cite[Theorem~A.4.1]{meyn_control_2008}, the system returns to $\s$ in finite mean time.

    We call a period of time inside the bounded set $\s$ an \textit{$\s$-visit}. Each $\s$-visit has a finite mean time because there is a positive probability of having a lot of arrivals in the next second and leaving $\s$. Moreover, as proved above using the Lyapunov argument, the time between two $\s$-visits has finite mean. 

    Each $\s$-visit has a positive probability of ending the $y^*$-cycle. To prove this, we construct a positive probability sample path of beginning a good period with $q^\msj = q^\ak$ and ending the good period in $(0,0,y^*,y^*)$, while remaining in $\s$. 
    \begin{itemize}
        \item First, we have a lot of completions in the two systems, completely emptying both. $q^\msj = q^\ak = |y^\msj| = 0$. Next, $k$ jobs arrive. Now $q^\ak=k$ and $q^\msj=0$. During this time $y^\msj \neq y^\ak$.
        \item Then $k$ jobs complete in the Ak system, no jobs complete in the MSJ system, and the newly generated Ak jobs are sampled such that $y^\msj = y^\ak$, while $q^\msj=q^\ak=0$.
        \item Next, $k+1$ jobs arrive, and a good period begins.
        \item Finally, $k+1$ jobs complete in both systems, ending with $y^\msj=y^\ak=y^*$, and $q^\msj=q^\ak=0$. Now a $y^*$-cycle ends, and the next begins.
    \end{itemize}
    All of these events have strictly positive probability
    and are independent of each other,
    so their joint occurrence has strictly positive probability as well.
    Thus, the length of a $y^*$-cycle is bounded by a geometric number of $\s$-visits, each of which has finite mean time,
    completing the proof.
\end{proof}

%% file: appendix.tex
\section{Finiteness of $\Delta$, and the conditions for drift lemma}
\label{app:basic-results}
\begin{lemma}\label{lem:delta-exists}
    The relative completion function
    \begin{align*}
    \Delta_\pi(y_1,y_2) = \lim_{t \to \infty} \Ep[C_\pi(y_1, t) - C_\pi(y_2, t)]
    \end{align*}
    is well-defined and finite for any pair of states $y_1$ and $y_2$ of the service process $\pi$.
\end{lemma}

\begin{proof}
    Throughout this proof, we leave the subscript $\pi$ implicit.
    
    To characterize $\Ep[C(y_1, t) - C(y_2, t)]$, we construct a coupling between the two instances of the service process $\pi$, starting with initial states $y_1$ and $y_2$. We let the two chains transition independently when their states are different, and let them transition identically once their states become the same. Let $\tau$ be the time that the states of the two systems become the same. Because the two systems remain identical after $\tau$, for any $t \geq 0$, 
    \begin{align*}
        C(y_1, t) - C(y_2, t) &= C(y_1, \min(t,  \tau)) - C(y_2, \min(t, \tau)). 
    \end{align*}
    We assume that the system $\pi$ is irreducible.
    Because each system is irreducible, the joint Markov chain of two systems is also irreducible and $\tau < \infty$ almost surely. Therefore, 
    \[
        \lim_{t\to\infty} \Ep[C(y_1, t) - C(y_2, t)] =  \Ep[{C(y_1, \tau) - C(y_2, \tau)}].
    \]
    The RHS of the above equality is clearly finite.
\end{proof}

Now we show that for any Markov chain $\eta$, 
\begin{align}\tag{\ref{eq:generator-steady-zero}}
    \Ep[G^\eta \circ f(Q^\eta, Y^\eta)] = 0.
\end{align}

The lemma below is implied by \cite[Proposition~3]{glynn_bounding_2008}:
\begin{replemma}{lem:drift-lemma}
    Let $f$ be a real-valued function of the state of a Markov chain $\eta$. Assume that the transition rates of the Markov chain $\eta$ are uniformly bounded, and $\Ep[f(Q^\eta, Y^\eta)] < \infty.$
    Then 
    \begin{equation}\tag{\ref{eq:generator-steady-zero}}
        \Ep_{(q,y) \sim (Q^\eta, Y^\eta)} [G^\eta \circ f(q, y)] = 0.
    \end{equation}
\end{replemma}

To check that the conditions of Lemma~\ref{lem:drift-lemma} hold for the At-least-$k$ and MSJ systems, first notice that their transitions rates are both uniformly bounded. In particular, the transition rates of the At-least-$k$ system are uniformly bounded by 
$\lambda + \max_y \sum_{y', a}\mu^\ak_{y,y',a}$, 
and the transition rates of the MSJ system are uniformly bounded by 
$\lambda + \max_{y, b} \sum_{y', a}\mu^\msj_{y,y',a, b}$.
Therefore we only need to check that each $f$ used in the paper has finite steady-state expectations in At-least-$k$ and MSJ systems, i.e. 
\begin{align*}
    \Ep[f(Q^\ak, Y^\ak)] &< \infty, \\
    \Ep[f(Q^\msj, Y^\msj)] &< \infty.
\end{align*}

The following lemma shows that a function $f$ has finite expectations in the At-least-$k$ and MSJ system as long as it grows at a polynomial rate in $q$, which is true for all $f$ which we will apply Lemma~\ref{lem:drift-lemma} to.
\begin{lemma}\label{lem:polynomial-finite-moments}
    Consider the MMSR system controlled by the Markov chain $\pi$ and the MSJ system. For any positive integer $m$, 
    \begin{align*}
        \Ep[(Q^\pi)^m] &< \infty, \\
        \Ep[(Q^\msj)^m] &< \infty.
    \end{align*}
\end{lemma}

To prove the lemma, we need \cite[Theorem~2.3]{hajek_hitting_1982}, restated as below:
\begin{lemma}  \label{lem:lyapunov-moment-bound}
    Consider a Markov chain $\eta$ with uniformly bounded total transition rates, and a Lyapunov function $V$ that satisfy the conditions below: $V(q, y) \geq 0$;
    there exists a constant $b, \gamma > 0$ such that whenever $V(q, y) \geq b$, 
    \begin{equation}\label{eq:lyapunov-moment-bound-neg-drift}
        G^\eta \circ V(q, y) \leq -\gamma;
    \end{equation}
    there exists $d> 0$ such that
    \begin{equation}\label{eq:lyapunov-moment-bound-bdd-jump}
        \max_{\text{next state } (q', y')} |V(q', y') - V(q, y)| \leq d.
    \end{equation}
    Then there exists $\theta > 0$ such that 
    \begin{equation}\label{eq:lyapunov-moment-bound-result}
        \Ep[e^{\theta V(Q^\eta, Y^\eta)}] < \infty.
    \end{equation}
\end{lemma}

Now we prove Lemma~\ref{lem:polynomial-finite-moments}.
\begin{proof}[Proof of Lemma~\ref{lem:polynomial-finite-moments}]
    We first prove the lemma for the MMSR system controlled by the Markov chain $\pi$. 
    
    Let $ \Delta_{\max}$ be the maximal absolute value of $\Delta_\pi(y)$ for any $y$ in the state spaces of $\yspi$, which must be finite due to Lemma~\ref{lem:delta-exists} and the fact that there are only finitely many possible $y$.

    We construct the Lyapunov function $V(q, y) = (q - \Delta(y))^+$. 
    We first check the conditions of Lemma~\ref{lem:lyapunov-moment-bound} for the MMSR system controlled by $\pi$. To check \eqref{eq:lyapunov-moment-bound-neg-drift}, we let $b = 1 + 2\Delta_{\max}$ and $\gamma = \comp - \lambda$. If $V(q, y) \geq b$, we must have $q\geq 1+\Delta_{\max}$; for any state $(q', y')$ that the system can jump to after one transition, $V(q', y') \geq q' - \Delta(y') \geq q - 1 - \Delta_{max} \geq 0$, so $V(q', y') = q' - \Delta(y')$. Therefore, 
    \[
        G^\pi \circ V(q, y) = G^\pi \circ (q - \Delta(y)) = \lambda - \comp = -\gamma.
    \]
    It is also easy to see that \eqref{eq:lyapunov-moment-bound-bdd-jump} holds with $d = 1 + 2\Delta_{\max}$. Therefore, by Lemma~\ref{lem:lyapunov-moment-bound}, there exists $\theta > 0$ such that
    \[
        \Ep[e^{\theta V(Q^\pi, Y^\pi)}] < \infty.
    \]
    Observe that $e^{\theta V(q, y)}$ grows with $q$ exponentially fast. Therefore, for any positive integer $m$, 
    \begin{align*}
        q^m &= O(e^{\theta V(q, y)}),  \\
        \Ep[(Q^\pi)^m] &< \infty.
    \end{align*}


    The analysis of the MSJ system is similar to the analysis of the At-least-$k$ system, which is a special case of the MMSR system with $\pi = \sat$. We consider the Lyapunov function 
    \[
        V(q, y) = 
        \begin{cases}
            \text{if } q > 1  & (q - \Delta_\sat(y))^+ \\
            \text{otherwise } & 0,
        \end{cases}
    \]
    and check the conditions of Lemma~\ref{lem:lyapunov-moment-bound}. Notice that $G^\msj \circ V(q, y) = G^\ak \circ V(q,y)$ for any $q \geq 1$, so the rest of the argument is verbatim. 
\end{proof}

%% file: main.bbl

\begin{thebibliography}{50}


\ifx \showCODEN    \undefined \def \showCODEN     #1{\unskip}     \fi
\ifx \showDOI      \undefined \def \showDOI       #1{#1}\fi
\ifx \showISBNx    \undefined \def \showISBNx     #1{\unskip}     \fi
\ifx \showISBNxiii \undefined \def \showISBNxiii  #1{\unskip}     \fi
\ifx \showISSN     \undefined \def \showISSN      #1{\unskip}     \fi
\ifx \showLCCN     \undefined \def \showLCCN      #1{\unskip}     \fi
\ifx \shownote     \undefined \def \shownote      #1{#1}          \fi
\ifx \showarticletitle \undefined \def \showarticletitle #1{#1}   \fi
\ifx \showURL      \undefined \def \showURL       {\relax}        \fi
\providecommand\bibfield[2]{#2}
\providecommand\bibinfo[2]{#2}
\providecommand\natexlab[1]{#1}
\providecommand\showeprint[2][]{arXiv:#2}

\bibitem[Afanaseva et~al\mbox{.}(2019)]%
        {afanaseva_stability_2019}
\bibfield{author}{\bibinfo{person}{Larisa Afanaseva}, \bibinfo{person}{Elena
  Bashtova}, {and} \bibinfo{person}{Svetlana Grishunina}.}
  \bibinfo{year}{2019}\natexlab{}.
\newblock \showarticletitle{Stability Analysis of a Multi-server Model with
  Simultaneous Service and a Regenerative Input Flow}.
\newblock \bibinfo{journal}{\emph{Methodology and Computing in Applied
  Probability}} (\bibinfo{year}{2019}), \bibinfo{pages}{1--17}.
\newblock


\bibitem[Baccelli and Foss(1995)]%
        {baccelli_1995}
\bibfield{author}{\bibinfo{person}{François Baccelli} {and}
  \bibinfo{person}{Serguei Foss}.} \bibinfo{year}{1995}\natexlab{}.
\newblock \showarticletitle{On the saturation rule for the stability of
  queues}.
\newblock \bibinfo{journal}{\emph{Journal of Applied Probability}}
  \bibinfo{volume}{32}, \bibinfo{number}{2} (\bibinfo{year}{1995}),
  \bibinfo{pages}{494–507}.
\newblock
\urldef\tempurl%
\url{https://doi.org/10.2307/3215303}
\showDOI{\tempurl}


\bibitem[Brill and Green(1984)]%
        {brill_queues_1984}
\bibfield{author}{\bibinfo{person}{Percy~H. Brill} {and} \bibinfo{person}{Linda
  Green}.} \bibinfo{year}{1984}\natexlab{}.
\newblock \showarticletitle{Queues in Which Customers Receive Simultaneous
  Service from a Random Number of Servers: A System Point Approach}.
\newblock \bibinfo{journal}{\emph{Management Science}} \bibinfo{volume}{30},
  \bibinfo{number}{1} (\bibinfo{year}{1984}), \bibinfo{pages}{51--68}.
\newblock


\bibitem[Carastan-Santos et~al\mbox{.}(2019)]%
        {carastan_one_2019}
\bibfield{author}{\bibinfo{person}{Danilo Carastan-Santos},
  \bibinfo{person}{Raphael~Y. De~Camargo}, \bibinfo{person}{Denis Trystram},
  {and} \bibinfo{person}{Salah Zrigui}.} \bibinfo{year}{2019}\natexlab{}.
\newblock \showarticletitle{One Can Only Gain by Replacing {EASY} Backfilling:
  A Simple Scheduling Policies Case Study}. In \bibinfo{booktitle}{\emph{2019
  19th IEEE/ACM International Symposium on Cluster, Cloud and Grid Computing
  (CCGRID)}}. \bibinfo{pages}{1--10}.
\newblock


\bibitem[Clarke(1956)]%
        {clarke_waiting_1956}
\bibfield{author}{\bibinfo{person}{A~Bruce Clarke}.}
  \bibinfo{year}{1956}\natexlab{}.
\newblock \showarticletitle{A waiting line process of Markov type}.
\newblock \bibinfo{journal}{\emph{The Annals of Mathematical Statistics}}
  (\bibinfo{year}{1956}), \bibinfo{pages}{452--459}.
\newblock


\bibitem[Delasay et~al\mbox{.}(2016)]%
        {delasay_modeling_2016}
\bibfield{author}{\bibinfo{person}{Mohammad Delasay}, \bibinfo{person}{Armann
  Ingolfsson}, {and} \bibinfo{person}{Bora Kolfal}.}
  \bibinfo{year}{2016}\natexlab{}.
\newblock \showarticletitle{Modeling Load and Overwork Effects in Queueing
  Systems with Adaptive Service Rates}.
\newblock \bibinfo{journal}{\emph{Operations Research}} \bibinfo{volume}{64},
  \bibinfo{number}{4} (\bibinfo{year}{2016}), \bibinfo{pages}{867--885}.
\newblock


\bibitem[Doroudi(2016)]%
        {doroudi_stochastic_2016}
\bibfield{author}{\bibinfo{person}{Sherwin Doroudi}.}
  \bibinfo{year}{2016}\natexlab{}.
\newblock \showarticletitle{Stochastic analysis of maintenance and routing
  policies in queueing systems}.
\newblock  (\bibinfo{year}{2016}).
\newblock


\bibitem[Eryilmaz and Srikant(2012)]%
        {eryilmaz_drift_2012}
\bibfield{author}{\bibinfo{person}{Atilla Eryilmaz} {and} \bibinfo{person}{R.
  Srikant}.} \bibinfo{year}{2012}\natexlab{}.
\newblock \showarticletitle{Asymptotically Tight Steady-State Queue Length
  Bounds Implied by Drift Conditions}.
\newblock \bibinfo{journal}{\emph{Queueing Syst. Theory Appl.}}
  \bibinfo{volume}{72}, \bibinfo{number}{3–4} (\bibinfo{date}{dec}
  \bibinfo{year}{2012}), \bibinfo{pages}{311–359}.
\newblock
\showISSN{0257-0130}
\urldef\tempurl%
\url{https://doi.org/10.1007/s11134-012-9305-y}
\showDOI{\tempurl}


\bibitem[Etsion and Tsafrir(2005)]%
        {etsion_short}
\bibfield{author}{\bibinfo{person}{Yoav Etsion} {and} \bibinfo{person}{Dan
  Tsafrir}.} \bibinfo{year}{2005}\natexlab{}.
\newblock \showarticletitle{A short survey of commercial cluster batch
  schedulers}.
\newblock \bibinfo{journal}{\emph{School of Computer Science and Engineering,
  The Hebrew University of Jerusalem}}  \bibinfo{volume}{44221}
  (\bibinfo{year}{2005}), \bibinfo{pages}{2005--13}.
\newblock


\bibitem[Feitelson et~al\mbox{.}(2004)]%
        {feitelson_parallel_2004}
\bibfield{author}{\bibinfo{person}{Dror~G. Feitelson}, \bibinfo{person}{Larry
  Rudolph}, {and} \bibinfo{person}{Uwe Schwiegelshohn}.}
  \bibinfo{year}{2004}\natexlab{}.
\newblock \showarticletitle{Parallel job scheduling—a status report}. In
  \bibinfo{booktitle}{\emph{Workshop on {Job} {Scheduling} {Strategies} for
  {Parallel} {Processing}}}. \bibinfo{publisher}{Springer},
  \bibinfo{address}{New York, NY, USA}, \bibinfo{pages}{1--16}.
\newblock


\bibitem[Filippopoulos and Karatza(2007)]%
        {fillippopoulos_mm2}
\bibfield{author}{\bibinfo{person}{Dimitrios Filippopoulos} {and}
  \bibinfo{person}{Helen Karatza}.} \bibinfo{year}{2007}\natexlab{}.
\newblock \showarticletitle{An M/M/2 parallel system model with pure space
  sharing among rigid jobs}.
\newblock \bibinfo{journal}{\emph{Mathematical and Computer Modelling}}
  \bibinfo{volume}{45}, \bibinfo{number}{5} (\bibinfo{year}{2007}),
  \bibinfo{pages}{491 -- 530}.
\newblock
\showISSN{0895-7177}


\bibitem[Foss and Konstantopoulos(2004)]%
        {foss_2004}
\bibfield{author}{\bibinfo{person}{Serguei Foss} {and} \bibinfo{person}{Takis
  Konstantopoulos}.} \bibinfo{year}{2004}\natexlab{}.
\newblock \showarticletitle{An overview of some stochastic stability methods}.
\newblock \bibinfo{journal}{\emph{Journal of the Operations Research Society of
  Japan}} \bibinfo{volume}{47}, \bibinfo{number}{4} (\bibinfo{year}{2004}),
  \bibinfo{pages}{275--303}.
\newblock


\bibitem[Ghaderi(2016)]%
        {ghaderi_randomized_2016}
\bibfield{author}{\bibinfo{person}{Javad Ghaderi}.}
  \bibinfo{year}{2016}\natexlab{}.
\newblock \showarticletitle{Randomized algorithms for scheduling {VMs} in the
  cloud}. In \bibinfo{booktitle}{\emph{{IEEE} {INFOCOM} 2016 - {The} 35th
  {Annual} {IEEE} {International} {Conference} on {Computer}
  {Communications}}}. \bibinfo{pages}{1--9}.
\newblock


\bibitem[Glynn et~al\mbox{.}(2008)]%
        {glynn_bounding_2008}
\bibfield{author}{\bibinfo{person}{Peter~W Glynn}, \bibinfo{person}{Assaf
  Zeevi}, {et~al\mbox{.}}} \bibinfo{year}{2008}\natexlab{}.
\newblock \showarticletitle{Bounding stationary expectations of Markov
  processes}.
\newblock \bibinfo{journal}{\emph{Markov processes and related topics: a
  Festschrift for Thomas G. Kurtz}}  \bibinfo{volume}{4}
  (\bibinfo{year}{2008}), \bibinfo{pages}{195--214}.
\newblock


\bibitem[Grosof and Harchol-Balter(2023)]%
        {grosof_serverfilling_2023}
\bibfield{author}{\bibinfo{person}{Isaac Grosof} {and} \bibinfo{person}{Mor
  Harchol-Balter}.} \bibinfo{year}{2023}\natexlab{}.
\newblock \showarticletitle{Invited Paper: {ServerFilling}: A Better Approach
  to Packing Multiserver Jobs}. In \bibinfo{booktitle}{\emph{Proceedings of the
  5th Workshop on Advanced Tools, Programming Languages, and PLatforms for
  Implementing and Evaluating Algorithms for Distributed Systems}} (Orlando,
  FL, USA) \emph{(\bibinfo{series}{ApPLIED 2023})}.
  \bibinfo{publisher}{Association for Computing Machinery},
  \bibinfo{address}{New York, NY, USA}, Article \bibinfo{articleno}{7},
  \bibinfo{numpages}{5}~pages.
\newblock
\showISBNx{9798400701283}
\urldef\tempurl%
\url{https://doi.org/10.1145/3584684.3597264}
\showDOI{\tempurl}


\bibitem[Grosof et~al\mbox{.}(2020)]%
        {grosof_stability_2020}
\bibfield{author}{\bibinfo{person}{Isaac Grosof}, \bibinfo{person}{Mor
  Harchol-Balter}, {and} \bibinfo{person}{Alan Scheller-Wolf}.}
  \bibinfo{year}{2020}\natexlab{}.
\newblock \showarticletitle{Stability for two-class multiserver-job systems}.
\newblock \bibinfo{journal}{\emph{arXiv preprint arXiv:2010.00631}}
  (\bibinfo{year}{2020}).
\newblock


\bibitem[Grosof et~al\mbox{.}(2022a)]%
        {grosof_wcfs_2021}
\bibfield{author}{\bibinfo{person}{Isaac Grosof}, \bibinfo{person}{Mor
  Harchol-Balter}, {and} \bibinfo{person}{Alan Scheller-Wolf}.}
  \bibinfo{year}{2022}\natexlab{a}.
\newblock \showarticletitle{{WCFS}: A new framework for analyzing multiserver
  systems}.
\newblock \bibinfo{journal}{\emph{Queueing Systems}} (\bibinfo{year}{2022}).
\newblock


\bibitem[Grosof et~al\mbox{.}(2023)]%
        {grosof_new_2023}
\bibfield{author}{\bibinfo{person}{Isaac Grosof}, \bibinfo{person}{Mor
  Harchol-Balter}, {and} \bibinfo{person}{Alan Scheller-Wolf}.}
  \bibinfo{year}{2023}\natexlab{}.
\newblock \showarticletitle{New stability results for multiserver-job models
  via product-form saturated systems}.
\newblock \bibinfo{journal}{\emph{MAthematical performance Modeling and
  Analysis (MAMA)}} \bibinfo{volume}{4}, \bibinfo{number}{6}
  (\bibinfo{year}{2023}), \bibinfo{pages}{1}.
\newblock


\bibitem[Grosof et~al\mbox{.}(2022b)]%
        {grosof_optimal_2022}
\bibfield{author}{\bibinfo{person}{Isaac Grosof}, \bibinfo{person}{Ziv Scully},
  \bibinfo{person}{Mor Harchol-Balter}, {and} \bibinfo{person}{Alan
  Scheller-Wolf}.} \bibinfo{year}{2022}\natexlab{b}.
\newblock \showarticletitle{Optimal Scheduling in the Multiserver-Job Model
  under Heavy Traffic}.
\newblock \bibinfo{journal}{\emph{Proc. ACM Meas. Anal. Comput. Syst.}}
  \bibinfo{volume}{6}, \bibinfo{number}{3}, Article \bibinfo{articleno}{51}
  (\bibinfo{date}{dec} \bibinfo{year}{2022}), \bibinfo{numpages}{32}~pages.
\newblock
\urldef\tempurl%
\url{https://doi.org/10.1145/3570612}
\showDOI{\tempurl}


\bibitem[Gupta et~al\mbox{.}(2006)]%
        {gupta_fundamental_2006}
\bibfield{author}{\bibinfo{person}{Varun Gupta}, \bibinfo{person}{Mor
  Harchol-Balter}, \bibinfo{person}{Alan~Scheller Wolf}, {and}
  \bibinfo{person}{Uri Yechiali}.} \bibinfo{year}{2006}\natexlab{}.
\newblock \showarticletitle{Fundamental characteristics of queues with
  fluctuating load}. In \bibinfo{booktitle}{\emph{Proceedings of the joint
  international conference on Measurement and modeling of computer systems}}.
  \bibinfo{pages}{203--215}.
\newblock


\bibitem[Hajek(1982)]%
        {hajek_hitting_1982}
\bibfield{author}{\bibinfo{person}{Bruce Hajek}.}
  \bibinfo{year}{1982}\natexlab{}.
\newblock \showarticletitle{Hitting-time and occupation-time bounds implied by
  drift analysis with applications}.
\newblock \bibinfo{journal}{\emph{Advances in Applied Probability}}
  \bibinfo{volume}{14}, \bibinfo{number}{3} (\bibinfo{year}{1982}),
  \bibinfo{pages}{502–525}.
\newblock
\urldef\tempurl%
\url{https://doi.org/10.2307/1426671}
\showDOI{\tempurl}


\bibitem[Hong(2022)]%
        {hong_sharp_2022}
\bibfield{author}{\bibinfo{person}{Yige Hong}.}
  \bibinfo{year}{2022}\natexlab{}.
\newblock \showarticletitle{Sharp Zero-Queueing Bounds for Multi-Server Jobs}.
\newblock \bibinfo{journal}{\emph{SIGMETRICS Perform. Eval. Rev.}}
  \bibinfo{volume}{49}, \bibinfo{number}{2} (\bibinfo{date}{jan}
  \bibinfo{year}{2022}), \bibinfo{pages}{66–68}.
\newblock
\showISSN{0163-5999}


\bibitem[Jones and Nitzberg(1999)]%
        {jones_scheduling}
\bibfield{author}{\bibinfo{person}{James~Patton Jones} {and}
  \bibinfo{person}{Bill Nitzberg}.} \bibinfo{year}{1999}\natexlab{}.
\newblock \showarticletitle{Scheduling for Parallel Supercomputing: A
  Historical Perspective of Achievable Utilization}. In
  \bibinfo{booktitle}{\emph{Job Scheduling Strategies for Parallel
  Processing}}, \bibfield{editor}{\bibinfo{person}{Dror~G. Feitelson} {and}
  \bibinfo{person}{Larry Rudolph}} (Eds.). \bibinfo{publisher}{Springer Berlin
  Heidelberg}, \bibinfo{address}{Berlin, Heidelberg}, \bibinfo{pages}{1--16}.
\newblock
\showISBNx{978-3-540-47954-3}


\bibitem[Knessl and Yang(2002)]%
        {knessl_exact_2002}
\bibfield{author}{\bibinfo{person}{Charles Knessl} {and}
  \bibinfo{person}{Yongzhi~Peter Yang}.} \bibinfo{year}{2002}\natexlab{}.
\newblock \showarticletitle{An exact solution for an M (t)/M (t)/1 queue with
  time-dependent arrivals and service}.
\newblock \bibinfo{journal}{\emph{Queueing systems}}  \bibinfo{volume}{40}
  (\bibinfo{year}{2002}), \bibinfo{pages}{233--245}.
\newblock


\bibitem[Lucantoni and Neuts(1994)]%
        {lucantoni_some_1994}
\bibfield{author}{\bibinfo{person}{David~M Lucantoni} {and}
  \bibinfo{person}{Marcel~F Neuts}.} \bibinfo{year}{1994}\natexlab{}.
\newblock \showarticletitle{Some steady-state distributions for the MAP/SM/1
  queue}.
\newblock \bibinfo{journal}{\emph{Stochastic Models}} \bibinfo{volume}{10},
  \bibinfo{number}{3} (\bibinfo{year}{1994}), \bibinfo{pages}{575--598}.
\newblock


\bibitem[Madni et~al\mbox{.}(2017)]%
        {madni_performance_2017}
\bibfield{author}{\bibinfo{person}{Syed Hamid~Hussain Madni},
  \bibinfo{person}{Muhammad~Shafie Abd~Latiff}, \bibinfo{person}{Mohammed
  Abdullahi}, \bibinfo{person}{Shafi'i~Muhammad Abdulhamid}, {and}
  \bibinfo{person}{Mohammed~Joda Usman}.} \bibinfo{year}{2017}\natexlab{}.
\newblock \showarticletitle{Performance comparison of heuristic algorithms for
  task scheduling in IaaS cloud computing environment}.
\newblock \bibinfo{journal}{\emph{PLOS ONE}} \bibinfo{volume}{12},
  \bibinfo{number}{5} (\bibinfo{date}{05} \bibinfo{year}{2017}),
  \bibinfo{pages}{1--26}.
\newblock
\urldef\tempurl%
\url{https://doi.org/10.1371/journal.pone.0176321}
\showDOI{\tempurl}


\bibitem[{Maguluri} and {Srikant}(2014)]%
        {maguluri_scheduling_2014}
\bibfield{author}{\bibinfo{person}{S.~T. {Maguluri}} {and} \bibinfo{person}{R.
  {Srikant}}.} \bibinfo{year}{2014}\natexlab{}.
\newblock \showarticletitle{Scheduling Jobs With Unknown Duration in Clouds}.
\newblock \bibinfo{journal}{\emph{IEEE/ACM Transactions on Networking}}
  \bibinfo{volume}{22}, \bibinfo{number}{6} (\bibinfo{year}{2014}),
  \bibinfo{pages}{1938--1951}.
\newblock


\bibitem[Maguluri and Srikant(2016)]%
        {Maguluri_drift_16}
\bibfield{author}{\bibinfo{person}{Siva~Theja Maguluri} {and}
  \bibinfo{person}{R. Srikant}.} \bibinfo{year}{2016}\natexlab{}.
\newblock \showarticletitle{Heavy traffic queue length behavior in a switch
  under the MaxWeight algorithm}.
\newblock  \bibinfo{volume}{6}, \bibinfo{number}{1} (\bibinfo{year}{2016}),
  \bibinfo{pages}{211--250}.
\newblock


\bibitem[Massey(1985)]%
        {massey_asymptotic_1985}
\bibfield{author}{\bibinfo{person}{William~A Massey}.}
  \bibinfo{year}{1985}\natexlab{}.
\newblock \showarticletitle{Asymptotic analysis of the time dependent M/M/1
  queue}.
\newblock \bibinfo{journal}{\emph{Mathematics of Operations Research}}
  \bibinfo{volume}{10}, \bibinfo{number}{2} (\bibinfo{year}{1985}),
  \bibinfo{pages}{305--327}.
\newblock


\bibitem[Meyn(2008)]%
        {meyn_control_2008}
\bibfield{author}{\bibinfo{person}{Sean Meyn}.}
  \bibinfo{year}{2008}\natexlab{}.
\newblock \bibinfo{booktitle}{\emph{Control techniques for complex networks}}.
\newblock \bibinfo{publisher}{Cambridge University Press}.
\newblock


\bibitem[Mitrani and Chakka(1995)]%
        {mitrani_spectral_1995}
\bibfield{author}{\bibinfo{person}{Isi Mitrani} {and} \bibinfo{person}{Ram
  Chakka}.} \bibinfo{year}{1995}\natexlab{}.
\newblock \showarticletitle{Spectral expansion solution for a class of Markov
  models: Application and comparison with the matrix-geometric method}.
\newblock \bibinfo{journal}{\emph{Performance Evaluation}}
  \bibinfo{volume}{23}, \bibinfo{number}{3} (\bibinfo{year}{1995}),
  \bibinfo{pages}{241--260}.
\newblock


\bibitem[Morozov and Rumyantsev(2016)]%
        {morozov_stability_2016}
\bibfield{author}{\bibinfo{person}{Evsey Morozov} {and}
  \bibinfo{person}{Alexander Rumyantsev}.} \bibinfo{year}{2016}\natexlab{}.
\newblock \showarticletitle{Stability Analysis of a MAP/M/s Cluster Model by
  Matrix-Analytic Method}. In \bibinfo{booktitle}{\emph{Computer Performance
  Engineering}}, \bibfield{editor}{\bibinfo{person}{Dieter Fiems},
  \bibinfo{person}{Marco Paolieri}, {and} \bibinfo{person}{Agapios~N. Platis}}
  (Eds.). \bibinfo{publisher}{Springer International Publishing},
  \bibinfo{address}{Cham}, \bibinfo{pages}{63--76}.
\newblock
\showISBNx{978-3-319-46433-6}


\bibitem[Neuts(1966)]%
        {neuts_single_1966}
\bibfield{author}{\bibinfo{person}{Marcel~F Neuts}.}
  \bibinfo{year}{1966}\natexlab{}.
\newblock \showarticletitle{The single server queue with Poisson input and
  semi-Markov service times}.
\newblock \bibinfo{journal}{\emph{Journal of Applied Probability}}
  \bibinfo{volume}{3}, \bibinfo{number}{1} (\bibinfo{year}{1966}),
  \bibinfo{pages}{202--230}.
\newblock


\bibitem[Newell(1968a)]%
        {newell_queuesIII_1968}
\bibfield{author}{\bibinfo{person}{GF Newell}.}
  \bibinfo{year}{1968}\natexlab{a}.
\newblock \showarticletitle{Queues with time-dependent arrival rates. III—A
  mild rush hour}.
\newblock \bibinfo{journal}{\emph{Journal of Applied Probability}}
  \bibinfo{volume}{5}, \bibinfo{number}{3} (\bibinfo{year}{1968}),
  \bibinfo{pages}{591--606}.
\newblock


\bibitem[Newell(1968b)]%
        {newell_queuesII_1968}
\bibfield{author}{\bibinfo{person}{GF Newell}.}
  \bibinfo{year}{1968}\natexlab{b}.
\newblock \showarticletitle{Queues with time-dependent arrival rates. II—The
  maximum queue and the return to equilibrium}.
\newblock \bibinfo{journal}{\emph{Journal of Applied Probability}}
  \bibinfo{volume}{5}, \bibinfo{number}{3} (\bibinfo{year}{1968}),
  \bibinfo{pages}{579--590}.
\newblock


\bibitem[Newell(1968c)]%
        {newell_queuesI_1968}
\bibfield{author}{\bibinfo{person}{Gordon~Frank Newell}.}
  \bibinfo{year}{1968}\natexlab{c}.
\newblock \showarticletitle{Queues with time-dependent arrival rates I—the
  transition through saturation}.
\newblock \bibinfo{journal}{\emph{Journal of Applied Probability}}
  \bibinfo{volume}{5}, \bibinfo{number}{2} (\bibinfo{year}{1968}),
  \bibinfo{pages}{436--451}.
\newblock


\bibitem[Peng(2022)]%
        {peng_exact_2022}
\bibfield{author}{\bibinfo{person}{Edwin Peng}.}
  \bibinfo{year}{2022}\natexlab{}.
\newblock \showarticletitle{Exact Response Time Analysis of Preemptive Priority
  Scheduling with Switching Overhead}.
\newblock \bibinfo{journal}{\emph{ACM SIGMETRICS Performance Evaluation
  Review}} \bibinfo{volume}{49}, \bibinfo{number}{2} (\bibinfo{year}{2022}),
  \bibinfo{pages}{72--74}.
\newblock


\bibitem[Perel and Yechiali(2008)]%
        {perel_queues_2008}
\bibfield{author}{\bibinfo{person}{Efrat Perel} {and} \bibinfo{person}{Uri
  Yechiali}.} \bibinfo{year}{2008}\natexlab{}.
\newblock \showarticletitle{Queues where customers of one queue act as servers
  of the other queue}.
\newblock \bibinfo{journal}{\emph{Queueing Systems}}  \bibinfo{volume}{60}
  (\bibinfo{year}{2008}), \bibinfo{pages}{271--288}.
\newblock


\bibitem[Psychas and Ghaderi(2018)]%
        {psychas_randomized_2018}
\bibfield{author}{\bibinfo{person}{Konstantinos Psychas} {and}
  \bibinfo{person}{Javad Ghaderi}.} \bibinfo{year}{2018}\natexlab{}.
\newblock \showarticletitle{Randomized Algorithms for Scheduling Multi-Resource
  Jobs in the Cloud}.
\newblock \bibinfo{journal}{\emph{IEEE/ACM Transactions on Networking}}
  \bibinfo{volume}{26}, \bibinfo{number}{5} (\bibinfo{year}{2018}),
  \bibinfo{pages}{2202--2215}.
\newblock


\bibitem[Rumyantsev(2020)]%
        {rumyantsev_stability_2020}
\bibfield{author}{\bibinfo{person}{Alexander Rumyantsev}.}
  \bibinfo{year}{2020}\natexlab{}.
\newblock \showarticletitle{Stability of multiclass multiserver models with
  automata-type phase transitions}. In \bibinfo{booktitle}{\emph{Proceedings of
  the second international workshop on stochastic modeling and applied research
  of technology (SMARTY 2020)}}, Vol.~\bibinfo{volume}{2792}.
  \bibinfo{pages}{213--225}.
\newblock


\bibitem[Rumyantsev et~al\mbox{.}(2022)]%
        {rumyantsev_three_2022}
\bibfield{author}{\bibinfo{person}{Alexander Rumyantsev},
  \bibinfo{person}{Robert Basmadjian}, \bibinfo{person}{Sergey Astafiev}, {and}
  \bibinfo{person}{Alexander Golovin}.} \bibinfo{year}{2022}\natexlab{}.
\newblock \showarticletitle{Three-level modeling of a speed-scaling
  supercomputer}.
\newblock \bibinfo{journal}{\emph{Annals of Operations Research}}
  (\bibinfo{year}{2022}), \bibinfo{pages}{1--29}.
\newblock


\bibitem[Rumyantsev and Morozov(2017)]%
        {rumyantsev_2017}
\bibfield{author}{\bibinfo{person}{Alexander Rumyantsev} {and}
  \bibinfo{person}{Evsey Morozov}.} \bibinfo{year}{2017}\natexlab{}.
\newblock \showarticletitle{Stability criterion of a multiserver model with
  simultaneous service}.
\newblock \bibinfo{journal}{\emph{Annals of Operations Research}}
  \bibinfo{volume}{252}, \bibinfo{number}{1} (\bibinfo{year}{2017}),
  \bibinfo{pages}{29--39}.
\newblock


\bibitem[Sliwko(2019)]%
        {sliwko_taxonomy_2019}
\bibfield{author}{\bibinfo{person}{Leszek Sliwko}.}
  \bibinfo{year}{2019}\natexlab{}.
\newblock \showarticletitle{A Taxonomy of Schedulers--Operating Systems,
  Clusters and Big Data Frameworks}.
\newblock \bibinfo{journal}{\emph{Global Journal of Computer Science and
  Technology}} (\bibinfo{year}{2019}).
\newblock


\bibitem[Srikant and Ying(2013)]%
        {srikant_communication_2013}
\bibfield{author}{\bibinfo{person}{Rayadurgam Srikant} {and}
  \bibinfo{person}{Lei Ying}.} \bibinfo{year}{2013}\natexlab{}.
\newblock \bibinfo{booktitle}{\emph{Communication networks: an optimization,
  control, and stochastic networks perspective}}.
\newblock \bibinfo{publisher}{Cambridge University Press}.
\newblock


\bibitem[Srinivasan et~al\mbox{.}(2002)]%
        {srinivasan_characterization_2002}
\bibfield{author}{\bibinfo{person}{Srividya Srinivasan},
  \bibinfo{person}{Rajkumar Kettimuthu}, \bibinfo{person}{Vijay Subramani},
  {and} \bibinfo{person}{Ponnuswamy Sadayappan}.}
  \bibinfo{year}{2002}\natexlab{}.
\newblock \showarticletitle{Characterization of backfilling strategies for
  parallel job scheduling}. In \bibinfo{booktitle}{\emph{Proceedings.
  International Conference on Parallel Processing Workshop}}.
  \bibinfo{pages}{514--519}.
\newblock


\bibitem[Tirmazi et~al\mbox{.}(2020)]%
        {tirmazi_2020}
\bibfield{author}{\bibinfo{person}{Muhammad Tirmazi}, \bibinfo{person}{Adam
  Barker}, \bibinfo{person}{Nan Deng}, \bibinfo{person}{Md~E. Haque},
  \bibinfo{person}{Zhijing~Gene Qin}, \bibinfo{person}{Steven Hand},
  \bibinfo{person}{Mor Harchol-Balter}, {and} \bibinfo{person}{John Wilkes}.}
  \bibinfo{year}{2020}\natexlab{}.
\newblock \showarticletitle{Borg: The next Generation}. In
  \bibinfo{booktitle}{\emph{Proceedings of the Fifteenth European Conference on
  Computer Systems}} (Heraklion, Greece) \emph{(\bibinfo{series}{EuroSys
  '20})}. \bibinfo{publisher}{Association for Computing Machinery},
  \bibinfo{address}{New York, NY, USA}, Article \bibinfo{articleno}{30},
  \bibinfo{numpages}{14}~pages.
\newblock
\showISBNx{9781450368827}


\bibitem[Vesilo et~al\mbox{.}(2022)]%
        {vesilo_scaling_2022}
\bibfield{author}{\bibinfo{person}{Rein Vesilo}, \bibinfo{person}{Mor
  Harchol-Balter}, {and} \bibinfo{person}{Alan Scheller-Wolf}.}
  \bibinfo{year}{2022}\natexlab{}.
\newblock \showarticletitle{Scaling properties of queues with time-varying load
  processes: extensions and applications}.
\newblock \bibinfo{journal}{\emph{Probability in the Engineering and
  Informational Sciences}} \bibinfo{volume}{36}, \bibinfo{number}{3}
  (\bibinfo{year}{2022}), \bibinfo{pages}{690--731}.
\newblock


\bibitem[Wang and Guo(2009)]%
        {wang_application_2009}
\bibfield{author}{\bibinfo{person}{Juan Wang} {and} \bibinfo{person}{Wenming
  Guo}.} \bibinfo{year}{2009}\natexlab{}.
\newblock \showarticletitle{The Application of Backfilling in Cluster Systems}.
  In \bibinfo{booktitle}{\emph{2009 WRI International Conference on
  Communications and Mobile Computing}}, Vol.~\bibinfo{volume}{3}.
  \bibinfo{pages}{55--59}.
\newblock


\bibitem[Wang et~al\mbox{.}(2021)]%
        {wang_zero_2021}
\bibfield{author}{\bibinfo{person}{Weina Wang}, \bibinfo{person}{Qiaomin Xie},
  {and} \bibinfo{person}{Mor Harchol-Balter}.} \bibinfo{year}{2021}\natexlab{}.
\newblock \showarticletitle{Zero Queueing for Multi-Server Jobs}. In
  \bibinfo{booktitle}{\emph{Abstract Proceedings of the 2021 ACM SIGMETRICS /
  International Conference on Measurement and Modeling of Computer Systems}}
  (Virtual Event, China) \emph{(\bibinfo{series}{SIGMETRICS '21})}.
  \bibinfo{publisher}{Association for Computing Machinery},
  \bibinfo{address}{New York, NY, USA}, \bibinfo{pages}{13–14}.
\newblock
\showISBNx{9781450380720}


\bibitem[Yechiali and Naor(1971)]%
        {yechiali_queuing_1971}
\bibfield{author}{\bibinfo{person}{Ury Yechiali} {and} \bibinfo{person}{Pinhas
  Naor}.} \bibinfo{year}{1971}\natexlab{}.
\newblock \showarticletitle{Queuing problems with heterogeneous arrivals and
  service}.
\newblock \bibinfo{journal}{\emph{Operations Research}} \bibinfo{volume}{19},
  \bibinfo{number}{3} (\bibinfo{year}{1971}), \bibinfo{pages}{722--734}.
\newblock


\end{thebibliography}
